\documentclass[
  aps,
  prd,
  11pt,
  superscriptaddress,
  notitlepage,
  onecolumn,
  nofootinbib
]{revtex4-1}

\AtBeginDocument{
  \linespread{1.05}\selectfont
}

\usepackage[
  a4paper,
  left=2.3cm,
  right=2.3cm,
  top=2.5cm,
  bottom=2.5cm
]{geometry}

\usepackage{microtype}
\usepackage{graphicx}
\usepackage{orcidlink}
\usepackage{enumerate}
\usepackage{enumitem}
\usepackage{ulem}
\usepackage{amssymb}
\usepackage[svgnames,dvipsnames]{xcolor}
\usepackage{tikz}
\usepackage{mathtools}
\usepackage{ytableau}
\usepackage{soul}
\usetikzlibrary{arrows.meta,positioning,fit,calc}
\hypersetup{hidelinks} 
\definecolor{linkcolor}{rgb}{0.0,0.3,0.5}
\definecolor{mypurple}{RGB}{143,116,210}

\graphicspath{{Figs/}}
\newcommand{\hias}{School of Fundamental Physics and Mathematical Sciences, Hangzhou Institute for Advanced Study, University of Chinese Academy of Sciences, Hangzhou 310024, China}

\begin{document}

\title{History-Dependent Mode Selection in a Driven Holographic Superfluid}

\author{Bo-Xuan Ge\orcidlink{0000-0003-0738-3473}}
\email{bo-xuan.ge@ucas.ac.cn}
\affiliation{\hias}

\begin{abstract}
We study finite-rate mode selection among competing Landau-unstable modes in a holographic superfluid. The first mode to reach a prescribed amplitude need not be the instantaneously fastest-growing one. Time-dependent linear evolution on the driven homogeneous background closely reproduces the modal amplitudes obtained from full bulk evolution. Protocols with identical present driving conditions can reverse the relative ordering of two modes. For the controlled histories studied here, the change in their logarithmic amplitude ratio is well approximated by the excess accumulated differential QNM gain.
\end{abstract}

\maketitle

\section{Introduction}
\label{sec:introduction}

The Landau criterion sets a fundamental limit on
dissipationless superflow. Above a critical velocity,
excitations can lower the energy of a moving fluid
\cite{Landau:1941Superfluidity}. Holographic superfluids
provide a setting in which this instability can be
studied at strong coupling
\cite{Herzog:2008he,Arean:2010zw,Sonner:2010yx,Arean:2010wu}.
Quasinormal-mode (QNM) calculations have shown that
the Landau threshold is accompanied by a dynamical
instability of counter-propagating excitations
\cite{Amado:2013aea}. Its relation to thermodynamic
instability has been examined in relativistic
superfluid hydrodynamics and holographic models
\cite{Gouteraux:2022kma,Arean:2023ejh}, while
nonlinear simulations have followed the decay of
unstable superflows through the formation of
solitonic structures \cite{Lan:2020kwn}.

QNM spectra determine the poles of retarded
correlation functions and characterise linear
relaxation and instability in holographic systems
\cite{Kovtun:2005ev}. Their properties in holographic
superfluids have been studied near the
symmetry-breaking transition and at finite momentum
\cite{Amado:2009ts,Arean:2021tks}. When several
unstable modes are present, their competition can
influence the subsequent nonlinear dynamics
\cite{Li:2020ayr}. On a fixed background, the
unstable spectrum can also constrain the structures
formed during nonlinear evolution, as demonstrated
in a rotating holographic superfluid
\cite{Yan:2026RotatingShell}.

This connection becomes less direct when the
background changes with time. Holographic quenches
have shown that finite-rate driving can produce
delayed responses, critical slowing down and defect
formation
\cite{Basu:2012gg,Sonner:2014tca,Chesler:2014gya,
Natsuume:2017jmu}. The importance of accumulated
growth during passage through an instability is
also familiar from slowly driven dynamical systems
\cite{Mandel:1987SlowPassage,Baer:1989SlowHopf}.
A mode with the largest instantaneous growth rate
need not have accumulated the greatest amplification.
Predicting the outcome of a finite-rate instability
therefore requires more than the spectrum at the
observation time.

Even the accumulated stationary-QNM growth need not
provide a complete description. A driven background
generally departs from the stationary solution at
the instantaneous control parameter. At the same
time, perturbations need not remain aligned with
a single instantaneous QNM as the linear operator
evolves. These effects are absent from adiabatic
transport along the stationary background family.
Their influence on the competition between unstable
finite-momentum modes remains to be determined.

We investigate this problem in a probe-limit
holographic superfluid at fixed mean charge density.
The imposed superflow parameter is ramped through
a region of competing Landau instabilities. We seed
eight counter-propagating Fourier modes with equal
boundary amplitudes and define selection by the
first mode to reach a prescribed threshold. Direct
bulk evolution is compared with adiabatically
transported QNMs and with time-dependent linear
evolution on stationary and driven homogeneous
backgrounds.

Adiabatic QNM transport reproduces most of the
observed mode-selection pattern, but fails in a
closely contested case at the fastest ramp and
largest threshold. There, the full evolution
selects $k=-0.875$, whereas stationary-QNM
transport predicts $k=-0.750$. Linear evolution
on the actual driven homogeneous background
reproduces the full result to high accuracy.
The discrepancy can therefore be accounted for
within linear perturbation dynamics, without
invoking nonlinear coupling between different
Fourier sectors.

To understand this discrepancy, we examine the
homogeneous response separately. The driven
background departs from the stationary family,
and its relaxation after the ramp is consistent
with the slowest physical $k=0$ QNM. We construct
a forced linear response about the moving
stationary family, including the finite-rate
boundary forcing. The reconstructed background
closely follows the homogeneous evolution and,
when used in the finite-momentum tangent equations,
recovers the observed selection in the tested
competing sectors. This identifies a contribution
from the background response that is absent from
stationary-QNM transport.

The comparison also reveals a partial cancellation
between the corrections associated with unrestricted
radial perturbation dynamics and the driven
homogeneous background. These corrections shift
the relative modal amplification in opposite
directions. Adiabatic transport can consequently
lie closer to the full evolution than linear
evolution on the instantaneous stationary
background, despite making a stronger approximation.
This behaviour is specific to the tested regime
and does not establish a general advantage of
adiabatic transport.

We further examine protocols with different
earlier histories but identical present driving
conditions. Their relative modal amplitudes can
have opposite orderings after the imposed drives
rejoin. For the controlled histories studied here,
the resulting change in the logarithmic amplitude
ratio of two modes is well approximated by their
excess accumulated differential QNM gain. This
provides a reduced description of pairwise modal
memory, distinct from the prediction of absolute
finite-rate mode selection.

Our analysis concerns the early amplification of
specified weak perturbations. The first-to-threshold
criterion does not determine the final nonlinear
state, and the accuracy of the forced-background
description is established only over the parameters
examined here.

The paper is organised as follows.
Section~\ref{sec:setup} introduces the holographic
superflow and its competing Landau-unstable modes.
Section~\ref{sec:selection} examines finite-rate
mode selection. Section~\ref{sec:tdlinear} studies
homogeneous background relaxation, forced response
and time-dependent linear amplification.
Section~\ref{sec:memory} considers the reversal
of pairwise modal ordering under different driving
histories, while Sec.~\ref{sec:gain} examines its
description through accumulated differential QNM gain.
Section~\ref{sec:conclusion} concludes. Numerical
methods and further validation are given in the
appendices.

\section{Holographic Superflow and Competing Landau Modes}
\label{sec:setup}

\subsection{Holographic superfluid}

We consider the Abelian--Higgs model in the probe limit,
\begin{equation}
    S_{\rm m}
    =
    \int d^4x\,\sqrt{-g}
    \left[
        -\frac14 F_{\mu\nu}F^{\mu\nu}
        -|D_\mu\Psi|^2
        -m^2|\Psi|^2
    \right],
    \label{eq:matter_action}
\end{equation}
where
\begin{equation}
    F_{\mu\nu}
    =
    \partial_\mu A_\nu-\partial_\nu A_\mu,
    \qquad
    D_\mu=\nabla_\mu-iA_\mu .
\end{equation}
After the standard probe-limit rescaling, we set the scalar charge to
unity and take $m^2=-2$.

The background is the planar Schwarzschild--AdS$_4$ black hole,
\begin{equation}
    ds^2
    =
    \frac{1}{z^2}
    \left[
        -f(z)\,dt_{\rm s}^2
        +\frac{dz^2}{f(z)}
        +dx^2+dy^2
    \right],
    \qquad
    f(z)=1-z^3 ,
    \label{eq:schwarzschild_ads}
\end{equation}
with the AdS boundary at $z=0$ and the horizon at $z=1$.
We set $L_{\rm AdS}=z_h=1$, so that
\begin{equation}
    T=T_H=\frac{3}{4\pi}.
\end{equation}

Near the AdS boundary,
\begin{equation}
    \Psi=z\Psi_-+z^2\Psi_++\cdots,\qquad A_t=\mu-\rho z+\cdots,\qquad A_x=a_x+j_x z+\cdots .
    \label{eq:boundary_expansion}
\end{equation}
We use the standard quantisation and impose
\begin{equation}
    \Psi_-=0 .
\end{equation}
The coefficient $\Psi_+$ is identified with the condensate in our
normalisation, while $\mu$ and $\rho$ denote the chemical potential
and charge density.

For a homogeneous condensate with constant boundary phase, the
gauge-invariant superfluid velocity reduces to
\begin{equation}
    v=-\frac{a_x}{\mu}.
    \label{eq:superfluid_velocity}
\end{equation}
Unless stated otherwise, we work at fixed charge density
\begin{equation}
    \rho=10.0417,
    \label{eq:rho_baseline}
\end{equation}
which, by the planar scaling symmetry, corresponds to
$T/T_c\simeq0.637$.

\subsection{Stationary superflow and quasinormal modes}

A stationary homogeneous superflow can be written as
\begin{equation}
    \Psi=z\,\phi(z),
    \qquad
    A=A_t(z)\,dt_{\rm s}+A_x(z)\,dx ,
    \label{eq:static_ansatz}
\end{equation}
where $\phi$ is real. The equations of motion reduce to
\begin{align}
    \partial_z\!\left(f\,\partial_z\phi\right)
    +
    \left(
        \frac{A_t^2}{f}
        -A_x^2-z
    \right)\phi
    &=0,
    \label{eq:static_phi}
    \\
    \partial_z^2 A_t
    -
    \frac{2\phi^2}{f}A_t
    &=0,
    \label{eq:static_at}
    \\
    \partial_z\!\left(f\,\partial_z A_x\right)
    -
    2\phi^2 A_x
    &=0.
    \label{eq:static_ax}
\end{align}

We construct the stationary branch in the canonical ensemble.
At the AdS boundary,
\begin{equation}
    \phi(0)=0,
    \qquad
    -A_t'(0)=\rho,
    \qquad
    A_x(0)+vA_t(0)=0 ,
    \label{eq:static_boundary_conditions}
\end{equation}
while the chemical potential is determined by the solution,
\begin{equation}
    \mu\equiv A_t(0).
\end{equation}
Regularity at the horizon gives
\begin{equation}
    A_t(1)=0,
\end{equation}
together with
\begin{equation}
    3\phi'(1)+[1+A_x(1)^2]\phi(1)=0,\qquad 3A_x'(1)+2\phi(1)^2A_x(1)=0.
    \label{eq:horizon_regular_static}
\end{equation}

For the perturbative and time-dependent calculations, we transform
to ingoing Eddington--Finkelstein coordinates,
\begin{equation}
    ds^2
    =
    \frac{1}{z^2}
    \left[
        -f(z)\,dt^2
        -2\,dt\,dz
        +dx^2+dy^2
    \right],
    \label{eq:ef_metric}
\end{equation}
and perform a gauge transformation restoring the radial gauge
\begin{equation}
    A_z=0.
\end{equation}
We define the rescaled complex scalar
\begin{equation}
    \psi\equiv\frac{\Psi}{z}.
    \label{eq:rescaled_scalar}
\end{equation}
The stationary background is then represented by
\begin{equation}
    \psi_0(z)
    =
    \phi(z)e^{i\vartheta(z)},
\end{equation}
where
\begin{equation}
    f\,\vartheta'(z)+A_t(z)=0 .
    \label{eq:ef_background_phase}
\end{equation}

We consider longitudinal perturbations of the form
\begin{equation}
    \delta X(t,z,x)
    =
    \delta X_k(z)e^{-i\omega t+ikx},
    \label{eq:qnm_ansatz}
\end{equation}
where
\begin{equation}
    \delta X_k
    =
    \left(
        \delta\psi_R,
        \delta\psi_I,
        \delta A_t,
        \delta A_x
    \right)_k .
\end{equation}
We take $v>0$, so that $k<0$ corresponds to
counter-propagating perturbations.

Linearisation gives the generalised eigenvalue problem
\begin{equation}
    \left[
        \mathcal A_k(v)
        +
        \omega\,\mathcal B_k(v)
    \right]
    \delta X_k=0 .
    \label{eq:qnm_generalized}
\end{equation}
The perturbations are regular at the horizon and source-free
at the AdS boundary,
\begin{equation}
    \delta\psi_R(0)
    =
    \delta\psi_I(0)
    =
    \delta A_t(0)
    =
    \delta A_x(0)
    =
    0.
    \label{eq:qnm_boundary}
\end{equation}

The fixed-$\rho$ condition specifies the stationary background
family. At each background, the QNM perturbations have vanishing
variations of all boundary sources. For non-zero momentum, the
spatially averaged charge perturbation vanishes, leaving the
total charge unchanged.

For fixed $v$ and $k$, Eq.~\eqref{eq:qnm_generalized} gives a
discrete set of quasinormal frequencies $\omega_\alpha(k;v)$,
where $\alpha$ labels the QNM branches. We follow the branch
that develops the Landau instability and denote its frequency
by $\omega_k(v)$. With the convention in
Eq.~\eqref{eq:qnm_ansatz}, its growth rate is
\begin{equation}
    \gamma_k(v)
    \equiv
    \operatorname{Im}\omega_k(v),
    \label{eq:growth_rate}
\end{equation}
so that $\gamma_k>0$ corresponds to a dynamical instability.

For the baseline background, the counter-propagating sound mode
becomes unstable in the hydrodynamic limit at
\begin{equation}
    v_{\rm L}\simeq0.40132,
    \label{eq:hydro_landau_threshold}
\end{equation}
consistent with the holographic Landau criterion
\cite{Amado:2013aea} and the benchmark of
Ref.~\cite{Lan:2020kwn}.

\subsection{Competing finite-momentum instabilities}

We impose periodic boundary conditions along $x$ with
\begin{equation}
    L_x=16\pi ,
\end{equation}
so that the allowed momenta are
\begin{equation}
    k_n
    =
    \frac{2\pi n}{L_x}
    =
    \frac{n}{8},
    \qquad
    n\in\mathbb Z .
    \label{eq:fourier_momenta}
\end{equation}
We seed the eight counter-propagating modes
\begin{equation}
    k_{-m}=-\frac{m}{8},
    \qquad
    m=1,\ldots,8.
    \label{eq:candidate_modes}
\end{equation}
These are the candidate modes used in the selection analysis.
They do not exhaust the Fourier modes allowed by the periodic
domain.

\begin{figure}[t]
    \centering
    \includegraphics[width=0.74\textwidth]
    {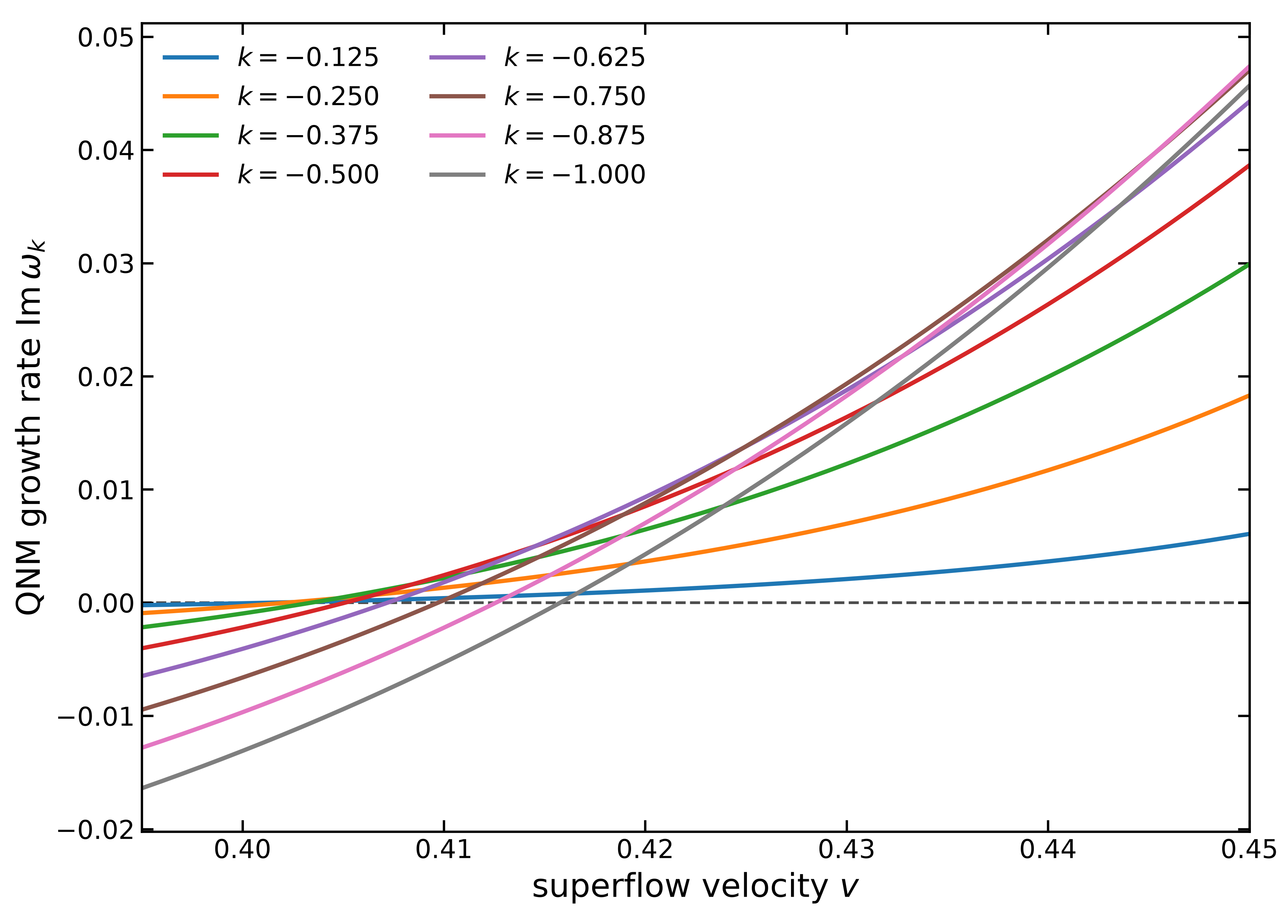}

    \vspace{0.6em}

    \includegraphics[width=0.74\textwidth]
    {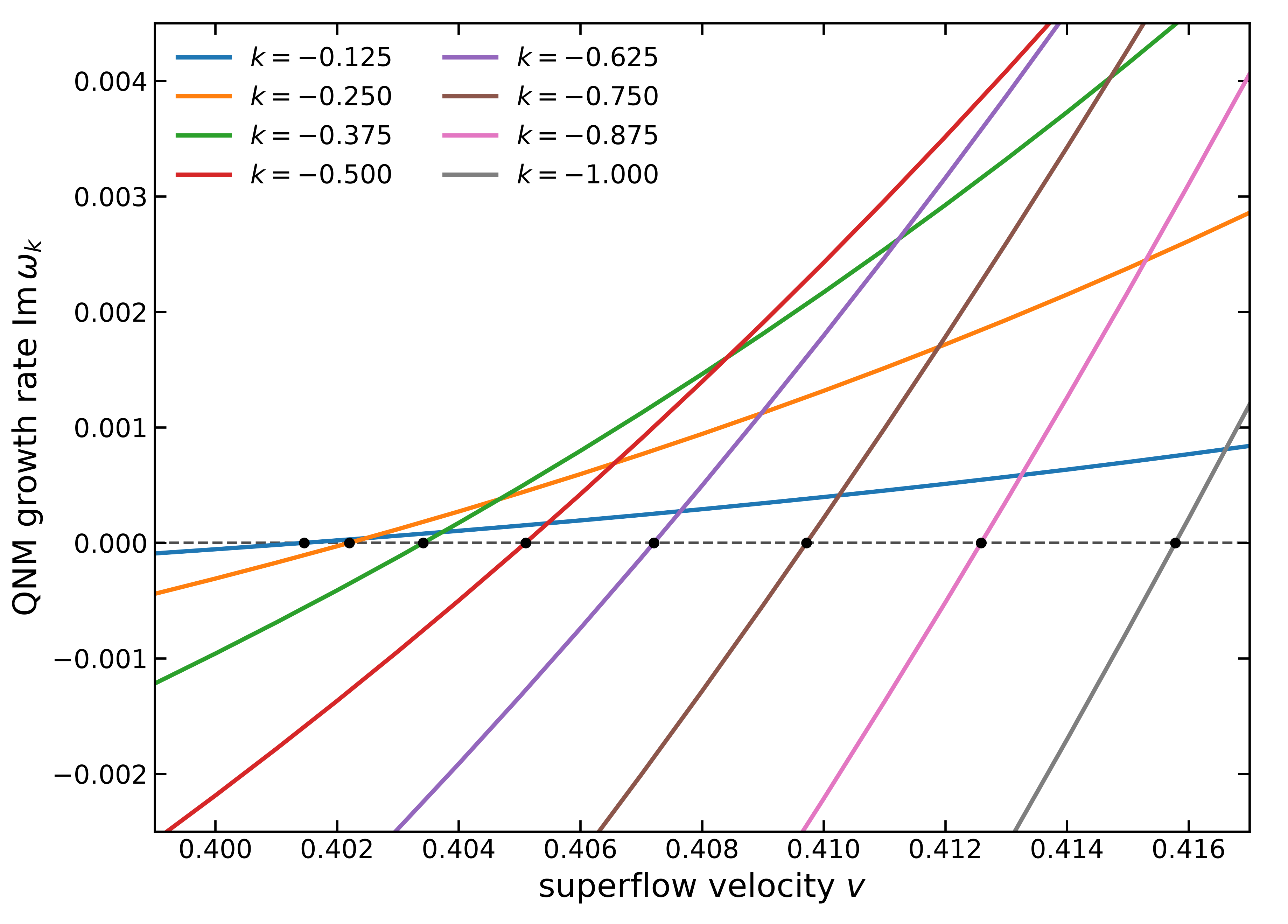}
    \caption{
        Growth rates $\gamma_k=\operatorname{Im}\omega_k$ of
        the eight counter-propagating QNM branches at
        $\rho=10.0417$.
        Upper: growth rates over $0.395\leq v\leq0.450$.
        Lower: the instability thresholds, with black dots
        marking the zero crossings $\gamma_k(v_c)=0$.
    }
    \label{fig:qnm_spectrum}
\end{figure}

Figure~\ref{fig:qnm_spectrum} shows their growth rates along
the stationary fixed-density branch. The finite-momentum
instability thresholds, defined by $\gamma_k(v_c)=0$, are
\begin{equation}
    \begin{gathered}
        v_c(-0.125)\simeq0.40146,\; v_c(-0.250)\simeq0.40220,\; v_c(-0.375)\simeq0.40342,\; v_c(-0.500)\simeq0.40510,\\
        v_c(-0.625)\simeq0.40721,\; v_c(-0.750)\simeq0.40972,\; v_c(-0.875)\simeq0.41259,\; v_c(-1.000)\simeq0.41578.
    \end{gathered}
    \label{eq:finite_k_thresholds}
\end{equation}
The values are obtained by linear interpolation of the
numerically determined QNM growth rates.

Modes with smaller $|k|$ become unstable first, but the
ordering of their growth rates changes as $v$ increases.
The first mode to become unstable therefore need not have
the largest growth rate later in the drive.

Finite-rate mode selection depends on both the onset of
instability and the subsequent amplification of the competing
modes. We examine this problem in the next section.

\section{Finite-Rate Mode Selection}
\label{sec:selection}

\subsection{Driving protocol and modal amplitudes}

We drive the system through the finite-momentum instability
region by varying the boundary sources along the stationary
family at fixed mean charge density,
\begin{equation}
    A_t(t,0)=\mu[v(t)],
    \qquad
    A_x(t,0)=-v(t)\mu[v(t)],
    \label{eq:driven_sources}
\end{equation}
where $v(t)$ is the imposed driving parameter. The bulk
fields evolve dynamically and need not follow the
instantaneous stationary solutions.

We consider linear ramps,
\begin{equation}
    v(t)=v_i+Rt,
    \qquad
    v_i=0.395,
    \qquad
    v_f=0.500,
    \label{eq:linear_ramp}
\end{equation}
with rates
\begin{equation}
    R=
    2.5\times10^{-4},\,
    4.0\times10^{-4},\,
    10^{-3}.
    \label{eq:ramp_rates}
\end{equation}
The initial velocity lies below the instability thresholds
of all eight candidate modes in
Eq.~\eqref{eq:candidate_modes}.

The local boundary charge density obeys
\begin{equation}
    \partial_t\rho(t,x)+\partial_x j_x(t,x)=0,
    \label{eq:boundary_charge_conservation}
\end{equation}
so periodicity preserves its spatial average,
\begin{equation}
    \bar\rho(t)
    \equiv
    \frac{1}{L_x}
    \int_0^{L_x}dx\,\rho(t,x)
    =
    10.0417 .
    \label{eq:mean_charge_conservation}
\end{equation}
The boundary sources thus follow the fixed-density
stationary family, although the evolving bulk
configuration need not remain on it.

The boundary condensate is obtained from the rescaled
scalar field,
\begin{equation}
    \mathcal O(t,x)
    \equiv
    \Psi_+(t,x)
    =
    \left.
    \partial_z\psi(t,z,x)
    \right|_{z=0}.
    \label{eq:boundary_condensate}
\end{equation}
Its Fourier coefficients and modal amplitudes are
\begin{equation}
    \mathcal O_k(t)
    =
    \frac{1}{L_x}
    \int_0^{L_x}dx\,
    \mathcal O(t,x)e^{-ikx},
    \qquad
    A_k(t)=|\mathcal O_k(t)|.
    \label{eq:modal_amplitude}
\end{equation}

The initial perturbation is a superposition of the eight
candidate QNMs, each normalised to the same one-sided
boundary Fourier amplitude,
\begin{equation}
    A_k(0)=10^{-5}.
    \label{eq:equal_mode_seed}
\end{equation}
This fixes the relative seed amplitudes independently of
the arbitrary normalisation of the QNM eigenvectors.

We define mode selection by the first crossing of a
prescribed amplitude threshold $A_{\rm th}$. For each
candidate mode, let $v_k(A_{\rm th};R)$ denote the driving
parameter at which its amplitude first reaches that
threshold. Among the modes that reach it, we define
\begin{equation}
    v_{\rm sel}(A_{\rm th};R)
    =
    \min_k v_k(A_{\rm th};R),
    \qquad
    k_{\rm sel}(A_{\rm th};R)
    =
    \operatorname*{arg\,min}_k
    v_k(A_{\rm th};R).
    \label{eq:selected_mode}
\end{equation}
We use three thresholds,
\begin{equation}
    A_{\rm th}
    =
    10^{-4},\,
    3\times10^{-4},\,
    10^{-3}.
    \label{eq:selection_thresholds}
\end{equation}
Here selection refers to the first of the eight seeded
modes to reach the specified amplitude, not to the final
state of the nonlinear evolution. The result is
conditional on the initial modal amplitudes and the
chosen threshold.

As a first comparison, we evaluate the stationary QNM
growth rates at the actual selection point and identify
the instantaneous fastest-growing mode,
\begin{equation}
    k_{\rm inst}(A_{\rm th};R)
    =
    \operatorname*{arg\,max}_k
    \gamma_k\!\left[
        v_{\rm sel}(A_{\rm th};R)
    \right].
    \label{eq:instantaneous_winner}
\end{equation}
This tests whether the mode reaching the threshold first
also has the largest instantaneous growth rate.

We also consider adiabatic transport of the stationary
QNM branches. For a linear ramp, the transported
amplitude is
\begin{equation}
    \ln A_k^{\rm tr}(v;R)
    =
    \ln A_k(0)
    +
    \frac{1}{R}
    \int_{v_i}^{v}
    \gamma_k(u)\,du
    +
    \ln
    \frac{\mathcal P_k(v)}
         {\mathcal P_k(v_i)},
    \label{eq:transported_qnm_amplitude}
\end{equation}
where $\mathcal P_k(v)>0$ accounts for the transport of
the QNM eigenvector and its projection onto the boundary
condensate observable. The construction is given in
Appendix~\ref{app:qnm_transport}.

Applying the same threshold criterion to $A_k^{\rm tr}$
defines the transported-QNM winner $k_{\rm sel}^{\rm tr}$.
Unlike the instantaneous criterion, this approximation
retains the growth accumulated along the stationary
background family. It follows a single instantaneous
QNM branch, however, and neglects both mixing with other
radial modes and the finite-rate departure of the
homogeneous background from the stationary family.

\subsection{Dynamical mode selection}

\begin{figure}[t]
    \centering
    \includegraphics[width=\textwidth]
    {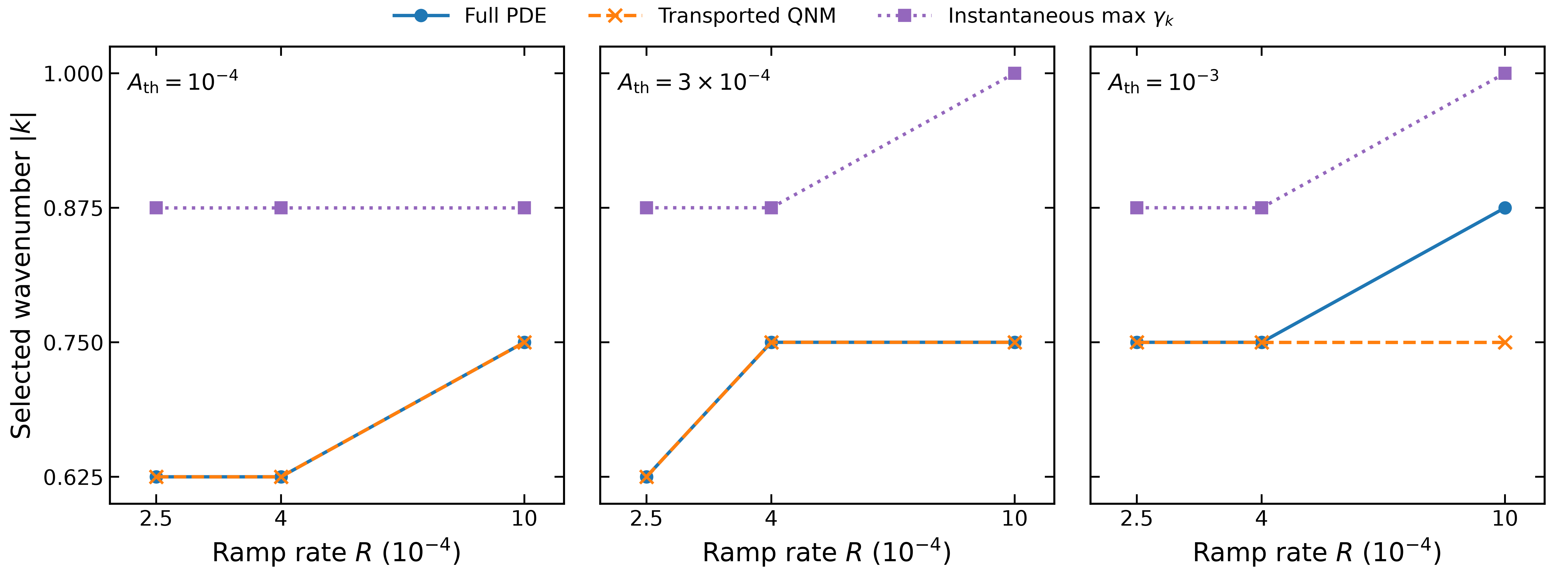}
    \caption{
        First-to-threshold mode selection among the eight
        seeded Fourier modes for three amplitude thresholds.
        The full bulk evolution is compared with the
        transported-QNM approximation and the mode having
        the largest instantaneous growth rate at the
        full-evolution selection point.
        The ramp rates are shown in units of $10^{-4}$.
        All candidate momenta are negative, so the
        vertical axis gives $|k|$.
        Lines connect the three sampled rates as a guide
        to the eye.
    }
    \label{fig:winner_map}
\end{figure}

Figure~\ref{fig:winner_map} compares the first-to-threshold
mode from the full bulk evolution with the transported-QNM
prediction and the instantaneous fastest-growing mode.
All eight candidates have negative momentum, so the
figure shows $|k|$.

At $A_{\rm th}=10^{-4}$, the full evolution selects
$|k|=0.625$ for $R=2.5\times10^{-4}$ and
$4.0\times10^{-4}$, but $|k|=0.750$ for $R=10^{-3}$.
The selected momentum therefore changes with the driving
rate, although the ramp endpoints and initial modal
amplitudes are unchanged.

For this threshold, the instantaneous criterion instead
selects $|k|=0.875$ at all three rates. The mode with the
largest growth rate at the selection point is therefore
not necessarily the first to reach the prescribed
amplitude.

The threshold also affects the selected mode. At
$R=2.5\times10^{-4}$, increasing $A_{\rm th}$ from
$10^{-4}$ to $10^{-3}$ changes the full-evolution
winner from $|k|=0.625$ to $0.750$. At $R=10^{-3}$,
the winner remains $|k|=0.750$ for the first two
thresholds, but changes to $0.875$ at
$A_{\rm th}=10^{-3}$.

Some selections are closely contested. At
$R=4.0\times10^{-4}$ and $A_{\rm th}=10^{-4}$,
the $k=-0.625$ and $k=-0.750$ modes reach the
threshold at velocities separated by only
$2.9\times10^{-6}$, as obtained from interpolation
of the full-evolution data.

The transported-QNM approximation reproduces the
full-evolution winner in eight of the nine combinations
of ramp rate and threshold shown in
Fig.~\ref{fig:winner_map}. The exception occurs at
\begin{equation}
    R=10^{-3},
    \qquad
    A_{\rm th}=10^{-3},
\end{equation}
where the full evolution selects $k=-0.875$, while
transported QNM predicts $k=-0.750$. The instantaneous
criterion selects $k=-1.000$.

For this case, the full evolution gives the first
threshold crossing at
\begin{equation}
    v_{\rm sel}^{\rm full}=0.49708898,
\end{equation}
for $k=-0.875$. The $k=-0.750$ mode reaches the same
threshold slightly later, at $v=0.49715691$. These
values are obtained by interpolation of the recorded
modal amplitudes.

The transported-QNM approximation captures most of
the finite-rate selection pattern, but its failure
in this case shows that stationary-family transport
is not always sufficient. To examine the missing
dynamical corrections, we first consider the response
of the homogeneous background to the finite-rate drive.

\section{Time-Dependent Linear Dynamics}
\label{sec:tdlinear}

The discrepancy found in Sec.~\ref{sec:selection} involves
two approximations in stationary-QNM transport. The
perturbation is restricted to a single instantaneous QNM
branch, and the homogeneous background is assumed to
follow the stationary superflow family. We examine these
approximations separately.

\subsection{Linear evolution on stationary and driven backgrounds}

Let $X_0(v)$ denote the stationary homogeneous background
and $\mathcal L_k[X]$ the linearised evolution operator
in Fourier sector $k$.

We first evolve the full radial perturbation on the
instantaneous stationary background,
\begin{equation}
    \partial_t\delta X_k
    =
    \mathcal L_k
    \left[
        X_0\bigl(v(t)\bigr)
    \right]\delta X_k .
    \label{eq:td_static}
\end{equation}
We call this TD-static. Unlike adiabatic PT, it allows
the perturbation to depart from the tracked Landau QNM
and mix with other radial modes. The background,
however, remains on the stationary family.

We then evolve an independent homogeneous background
$X_{\rm d}(t)$ under the same boundary sources as the
full calculation. The finite-momentum perturbations
satisfy
\begin{equation}
    \partial_t\delta X_k
    =
    \mathcal L_k
    \left[
        X_{\rm d}(t)
    \right]\delta X_k .
    \label{eq:td_driven}
\end{equation}
We refer to this calculation as TD-driven. The tangent
equations follow from the same bulk equations used in
the full evolution and are given explicitly in
Appendix~\ref{app:td_linear}.

Since the background is homogeneous, different Fourier
sectors remain decoupled at linear order. TD-driven
therefore includes the radial dynamics within each
momentum sector and the finite-rate response of the
homogeneous background, without nonlinear coupling
between distinct momenta.

\subsection{Homogeneous background relaxation}

During the ramp, the driven background generally
departs from the instantaneous stationary solution,
\begin{equation}
    X_{\rm d}(t)
    \neq
    X_0\bigl(v(t)\bigr).
    \label{eq:driven_background_lag}
\end{equation}
To quantify this departure without ambiguity from
the residual global scalar phase, we use the scalar
modulus and define
\begin{equation}
    \mathcal B_\psi(t)
    =
    \frac{
        \left\|
        |\psi_{\rm d}(t,z)|
        -
        |\psi_0(v(t),z)|
        \right\|_{L^2_z}
    }{
        \left\|
        |\psi_0(v(t),z)|
        \right\|_{L^2_z}
    }.
    \label{eq:homogeneous_lag}
\end{equation}

The lag grows with the driving rate. At $v=0.470$,
we find $\mathcal B_\psi\simeq1.687\times10^{-3}$
for $R=2.5\times10^{-4}$ and
$\mathcal B_\psi\simeq6.106\times10^{-3}$ for
$R=10^{-3}$. The departure is small but changes
the background on which the finite-momentum
perturbations evolve.

An independent measure of the homogeneous response
is provided by the physical $k=0$ QNMs. After
removing the global $U(1)$ phase zero mode, we
define the slowest relaxation time by
\begin{equation}
    \tau_0(v)
    =
    -\frac{1}{
        \operatorname{Re}\lambda_0(v)
    },
    \label{eq:k0_relaxation_time}
\end{equation}
where $\lambda_0$ is the least damped physical
eigenvalue of the homogeneous tangent operator.

Along the stationary branch, $\tau_0$ increases
from approximately $1.812$ at $v=0.420$ to
$4.913$ at $v=0.490$, reaching $7.307$ at
$v=0.500$.

To test this timescale, we hold the driving
parameter at $v=0.500$ after the ramp and
measure the subsequent decay of the homogeneous
background deviation. Table~\ref{tab:homogeneous_response}
compares the fitted decay times with the
stationary-QNM result.

\begin{table}[t]
    \centering
    \caption{
        Homogeneous response at $\rho=10.0417$.
        The scalar-modulus lag $\mathcal B_\psi$
        is evaluated at $v=0.470$.
        The decay times are fitted after the ramp
        ends at $v=0.500$, where the slowest
        physical $k=0$ QNM gives $\tau_0=7.307$.
    }
    \label{tab:homogeneous_response}
    \begin{tabular}{lcc}
        \hline
        $R$
        & $\mathcal B_\psi(v=0.470)$
        & $\tau_{\rm fit}$
        \\
        \hline
        $2.5\times10^{-4}$
        & $1.687\times10^{-3}$
        & $7.27$
        \\
        $4.0\times10^{-4}$
        & $2.643\times10^{-3}$
        & $7.25$
        \\
        $1.0\times10^{-3}$
        & $6.106\times10^{-3}$
        & $7.21$
        \\
        \hline
    \end{tabular}
\end{table}

The fitted decay times are close to the QNM
prediction. This agreement connects the
post-ramp relaxation to the homogeneous spectrum.
During the ramp, however, the background is
continuously forced and need not follow a single
decaying mode.

\subsection{Forced response about the stationary family}

We next test whether the homogeneous background
can be reconstructed from its linear response
about the moving stationary family.

Let $Y=(\psi,A_x)$ denote the independent
homogeneous fields, with $A_t$ reconstructed
from the radial Maxwell constraint. We write
the forced scalar and gauge fields as
\begin{equation}
    \psi_{\rm F}(t,z) = e^{i\Theta(t)} \left[ \psi_0(v(t),z)+\delta\psi(t,z) \right],\qquad A_{x,\rm F}(t,z) = A_{x,0}(v(t),z)+\delta A_x(t,z).
    \label{eq:forced_background_expansion}
\end{equation}
The phase $\Theta(t)$ accounts for the global
scalar-phase motion, while
$\delta Y=(\delta\psi,\delta A_x)$ describes
the departure in a phase-fixed frame.

Let $\mathcal F[Y;v,R]$ be the homogeneous
evolution operator, including contributions
from the time-dependent boundary sources.
To first order in the departure from the
stationary family, the forced response obeys
\begin{equation}
    \partial_t\delta Y = \mathcal J_0(v)\delta Y + \mathcal F[Y_0(v);v,R] - R\,\partial_vY_0(v) - \dot\Theta \begin{pmatrix} i\psi_0(v)\\ 0 \end{pmatrix},
    \label{eq:forced_background_response}
\end{equation}
where $\mathcal J_0$ is the homogeneous tangent
operator. Terms quadratic in the response
are neglected.

The phase is fixed by imposing
\begin{equation}
    \operatorname{Im}
    \int_0^1 dz\,
    \psi_0^*(v,z)\,\delta\psi(t,z)
    =
    0.
    \label{eq:forced_phase_condition}
\end{equation}
This removes the component along the global
phase zero mode.

The forcing contains both the motion of the
stationary family and the explicit rate-dependent
boundary-source terms. Keeping only
$-R\,\partial_vY_0$ would therefore omit part
of the forcing present in the original
homogeneous equations.

We integrate Eq.~\eqref{eq:forced_background_response}
and reconstruct the homogeneous background
$X_{\rm F}(t)$. At $v=0.470$, the resulting
scalar-modulus lag agrees with the independently
evolved homogeneous background to within
$0.3\%$ in relative terms for all three rates.
The forced linear response thus reproduces
the small departure from the stationary family
over the range tested.

To examine its modal content, we project the
evolving response onto the instantaneous
physical $k=0$ eigenmodes. The slowest mode
reproduces the scalar-modulus response well.
For $R=10^{-3}$ at $v=0.470$, removing its
contribution leaves a relative radial
reconstruction residual of approximately
$2.13\%$.

The radial boundary response of $A_x$ is less
accurately described by this mode alone. At the
same point, its projected contribution is about
$1.51$ times the full response in
$\partial_z A_x|_{z=0}$. Other radial modes
therefore provide a substantial compensating
contribution. The slowest QNM captures
most of the scalar lag, but the driven
background is not generally a one-mode
response.

Finally, we use the reconstructed background
in the finite-momentum tangent equations,
\begin{equation}
    \partial_t\delta X_k
    =
    \mathcal L_k
    \left[
        X_{\rm F}(t)
    \right]\delta X_k .
    \label{eq:td_forced}
\end{equation}
We refer to this as the forced-background
calculation. Unlike TD-driven, it does not
require the nonlinear homogeneous trajectory
as an input.

\subsection{Finite-momentum amplification}

We now compare the different descriptions
for the fastest ramp,
\begin{equation}
    R=10^{-3},
\end{equation}
where adiabatic PT fails to reproduce the
first-to-threshold winner at $A_{\rm th}=10^{-3}$.

\begin{figure}[t]
    \centering
    \includegraphics[width=0.75\textwidth]
    {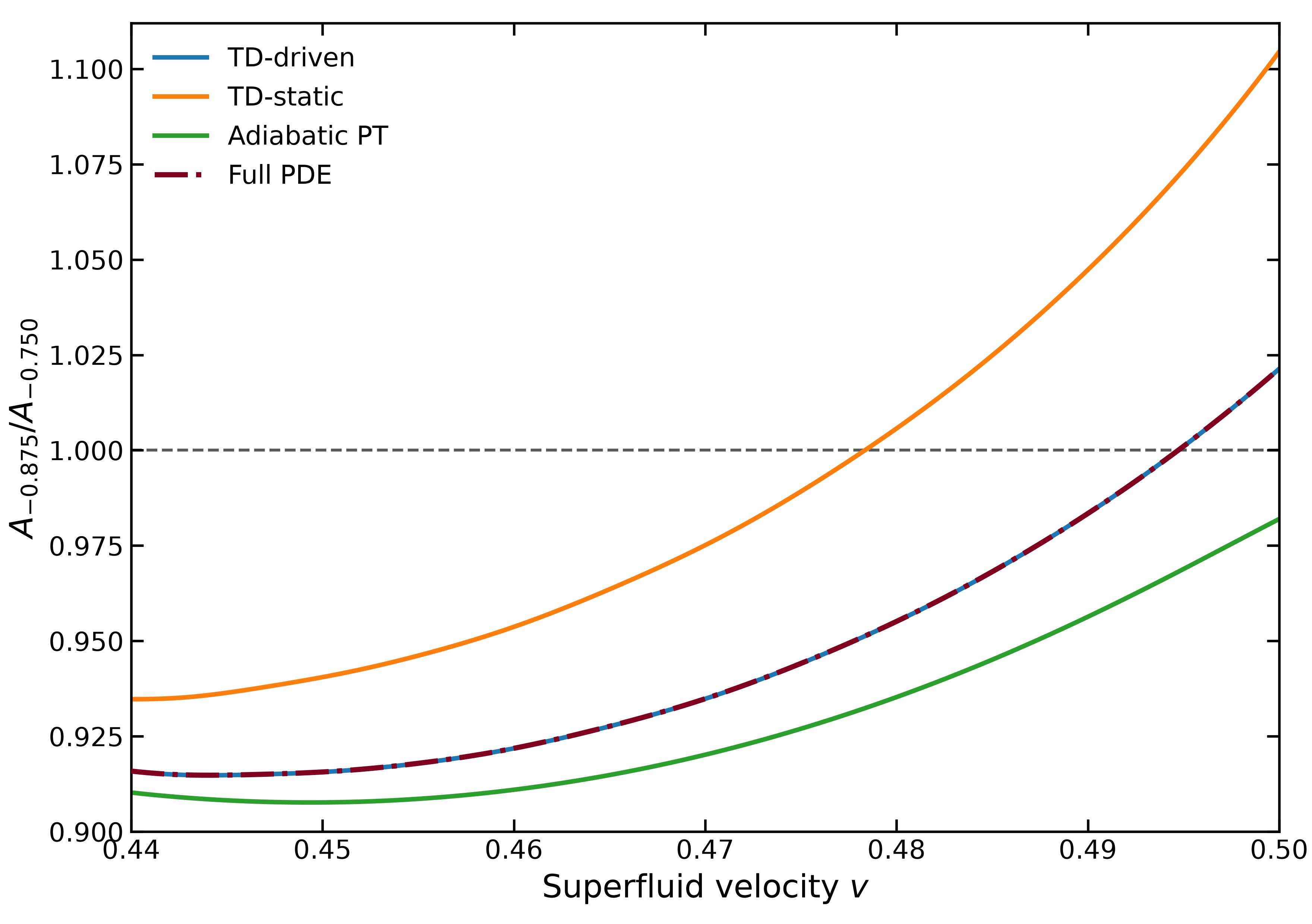}
    \caption{
        Relative amplitudes of the $k=-0.875$
        and $k=-0.750$ modes for $R=10^{-3}$.
        The full eight-mode bulk evolution
        (burgundy dashed curve) is compared
        with TD-driven, TD-static and adiabatic PT.
        The horizontal dashed line marks equal
        modal amplitudes.
        Full PDE and TD-driven nearly coincide,
        while TD-static crosses unity earlier
        and adiabatic PT remains below unity
        throughout the displayed interval.
    }
    \label{fig:td_driven}
\end{figure}

Figure~\ref{fig:td_driven} shows the ratio
\begin{equation}
    \mathcal R_{76}(v)
    =
    \frac{A_{-0.875}(v)}{A_{-0.750}(v)}
    \label{eq:fast_modal_ratio}
\end{equation}
from the full bulk evolution, TD-driven,
TD-static and adiabatic PT.

The full eight-mode evolution reaches equal
amplitudes at
\begin{equation}
    v_\times^{\rm full}
    =
    0.49472528,
\end{equation}
while TD-driven gives
\begin{equation}
    v_\times^{\rm TD\text{-}driven}
    =
    0.49472880.
    \label{eq:fast_crossing_td}
\end{equation}
The crossing positions differ by about
$3.52\times10^{-6}$.

TD-static predicts a considerably earlier
crossing,
\begin{equation}
    v_\times^{\rm TD\text{-}static}
    =
    0.47835051,
\end{equation}
whereas adiabatic PT does not reach equal
amplitudes before $v=0.500$.

The agreement between Full PDE and TD-driven
is consistent with the small perturbation
amplitudes used here. It shows that
nonlinear coupling between different Fourier
sectors is not needed to explain this
modal reordering.

The failure of TD-static is more informative:
allowing unrestricted radial linear dynamics
on the stationary family is not sufficient.
The response of the homogeneous background
must also be included.

To quantify the difference, we define
\begin{equation}
    \Delta_{76}(v)
    =
    \ln
    \frac{A_{-0.875}(v)}{A_{-0.750}(v)}.
    \label{eq:fast_log_modal_ratio}
\end{equation}
Table~\ref{tab:nonadiabatic_modal_ordering}
compares the models at $v=0.497$.
The forced-background entry comes from
Eq.~\eqref{eq:td_forced}, rather than from
the four curves displayed in
Fig.~\ref{fig:td_driven}.

\begin{table}[t]
    \centering
    \caption{
        Modal ordering at $v=0.497$ for
        $R=10^{-3}$, measured by
        $\Delta_{76}
        =\ln(A_{-0.875}/A_{-0.750})$.
        The forced-background result uses
        the reconstructed homogeneous response
        in the finite-momentum tangent equations.
    }
    \label{tab:nonadiabatic_modal_ordering}
    \begin{tabular}{lc}
        \hline
        Description & $\Delta_{76}$ \\
        \hline
        Adiabatic PT
        & $-0.026240$ \\
        TD-static
        & $+0.081812$ \\
        Forced-background
        & $+0.00864$ \\
        TD-driven
        & $+0.008767$ \\
        Full PDE
        & $+0.008783$ \\
        \hline
    \end{tabular}
\end{table}

The changes between the successive
approximations have opposite signs:
\begin{equation}
    \Delta_{76}^{\rm TD\text{-}static}-\Delta_{76}^{\rm PT}\simeq+0.10805,
    \qquad \Delta_{76}^{\rm forced}-\Delta_{76}^{\rm TD\text{-}static}\simeq-0.0732.
    \label{eq:nonadiabatic_corrections}
\end{equation}
The first difference measures the effect of
unrestricted radial linear dynamics on the
stationary background. The second measures
the change obtained by replacing that
background with the forced homogeneous
response.

These are differences between specified
approximations, not a unique separation
into independent physical processes.
Their partial cancellation explains why
adiabatic PT can be closer to the full
evolution than TD-static, despite making
a stronger approximation.

The forced-background result is much closer
to TD-driven. For $A_{\rm th}=10^{-3}$,
it selects $k=-0.875$, with the first
threshold crossing at
\begin{equation}
    v_{\rm sel}^{\rm forced}
    \simeq
    0.49709959.
\end{equation}
For comparison,
\begin{equation}
    v_{\rm sel}^{\rm TD\text{-}driven} = 0.49708914,\qquad v_{\rm sel}^{\rm full} = 0.49708898.
    \label{eq:fast_forced_selection}
\end{equation}
The forced response therefore recovers
the winner missed by adiabatic PT
without using the exact nonlinear
homogeneous trajectory.

The agreement also holds in the other
tested combinations of ramp rate and
threshold. These reduced finite-momentum
calculations include the modes competing
for the observed winner, rather than
repeating all eight linear sectors.
The result should therefore be understood
within the tested candidate sets and
thresholds.

Taken together, these comparisons show
that the finite-rate correction involves
both the radial dynamics of the unstable
perturbations and the response of the
homogeneous background. We next examine
how different driving histories can leave
distinct modal amplitudes even when the
imposed driving conditions become identical.

\section{History-Dependent Modal Ordering}
\label{sec:memory}

We now consider whether different driving histories can
leave different modal amplitudes at the same present
driving conditions. To isolate this effect, we compare
protocols that differ during an earlier dwell but follow
the same ramp afterwards.

These calculations use the original six-mode initial
perturbation, with equal boundary amplitudes of
$10^{-5}$. We examine the relative ordering of a
specified mode pair, rather than the first-to-threshold
winner among the eight candidates considered in
Sec.~\ref{sec:selection}.

\subsection{Same-present driving protocols}

We consider the pair
\begin{equation}
    k=-0.625,
    \qquad
    k=-0.750,
\end{equation}
whose stationary instability thresholds are
\begin{equation}
    v_c(-0.625)\simeq0.40721,
    \qquad
    v_c(-0.750)\simeq0.40972.
\end{equation}

Starting from $v_i=0.395$, we increase the imposed
superflow parameter at a rate
\begin{equation}
    R=4.25\times10^{-4}
\end{equation}
until it reaches
\begin{equation}
    v_h=0.4086.
    \label{eq:memory_hold_velocity}
\end{equation}
The drive is held at this value for a time $\tau_h$
before resuming at the original rate:
\begin{equation}
    v(t)
    =
    \begin{cases}
        v_i+Rt,
        & t<t_h,
        \\[2mm]
        v_h,
        & t_h\leq t<t_h+\tau_h,
        \\[2mm]
        v_h+R(t-t_h-\tau_h),
        & t\geq t_h+\tau_h,
    \end{cases}
    \qquad
    t_h=\frac{v_h-v_i}{R}.
    \label{eq:dwell_protocol}
\end{equation}

The hold lies between the two instability thresholds.
At $v_h$, the stationary QNM growth rates are
\begin{equation}
    \gamma_{-0.625}(v_h)
    =
    8.77602\times10^{-4},
    \qquad
    \gamma_{-0.750}(v_h)
    =
    -8.42512\times10^{-4}.
    \label{eq:memory_hold_growth}
\end{equation}
Thus the $k=-0.625$ mode grows during the dwell,
while the $k=-0.750$ mode is damped.

After the hold, all protocols follow the same ramp.
We compare them at
\begin{equation}
    v_{\rm obs}=0.460,
    \qquad
    \dot v_{\rm obs}=4.25\times10^{-4}.
    \label{eq:memory_observation_point}
\end{equation}
The observation times differ between protocols,
but the imposed boundary sources and their rates
of change are identical. The evolved bulk states
need not coincide.

\subsection{Reversal of pairwise modal ordering}

We define the logarithmic amplitude ratio at the
observation point by
\begin{equation}
    \Delta_{\rm mem}(\tau_h)
    =
    \ln
    \frac{
        A_{-0.750}(v_{\rm obs};\tau_h)
    }{
        A_{-0.625}(v_{\rm obs};\tau_h)
    }.
    \label{eq:memory_ratio}
\end{equation}
A positive value favours $k=-0.750$, whereas a
negative value favours $k=-0.625$.

Without a dwell, the full evolution gives
\begin{equation}
    \Delta_{\rm mem}(0)
    =
    4.20284\times10^{-2}.
\end{equation}
For dwell times $\tau_h=20$ and $30$, we obtain
\begin{equation}
    \Delta_{\rm mem}(20)=7.63547\times10^{-3},\qquad \Delta_{\rm mem}(30)=-9.58201\times10^{-3}.
    \label{eq:memory_reversal_values}
\end{equation}
The relative ordering reverses between these
histories, despite their identical present
driving conditions.

The TD-driven linear evolution gives
\begin{equation}
    \Delta_{\rm mem}^{\rm TD}(20) = 7.60340\times10^{-3},\qquad \Delta_{\rm mem}^{\rm TD}(30) = -9.61269\times10^{-3},
\end{equation}
closely following the full evolution.

\begin{figure}[t]
    \centering
    \includegraphics[width=\textwidth]
    {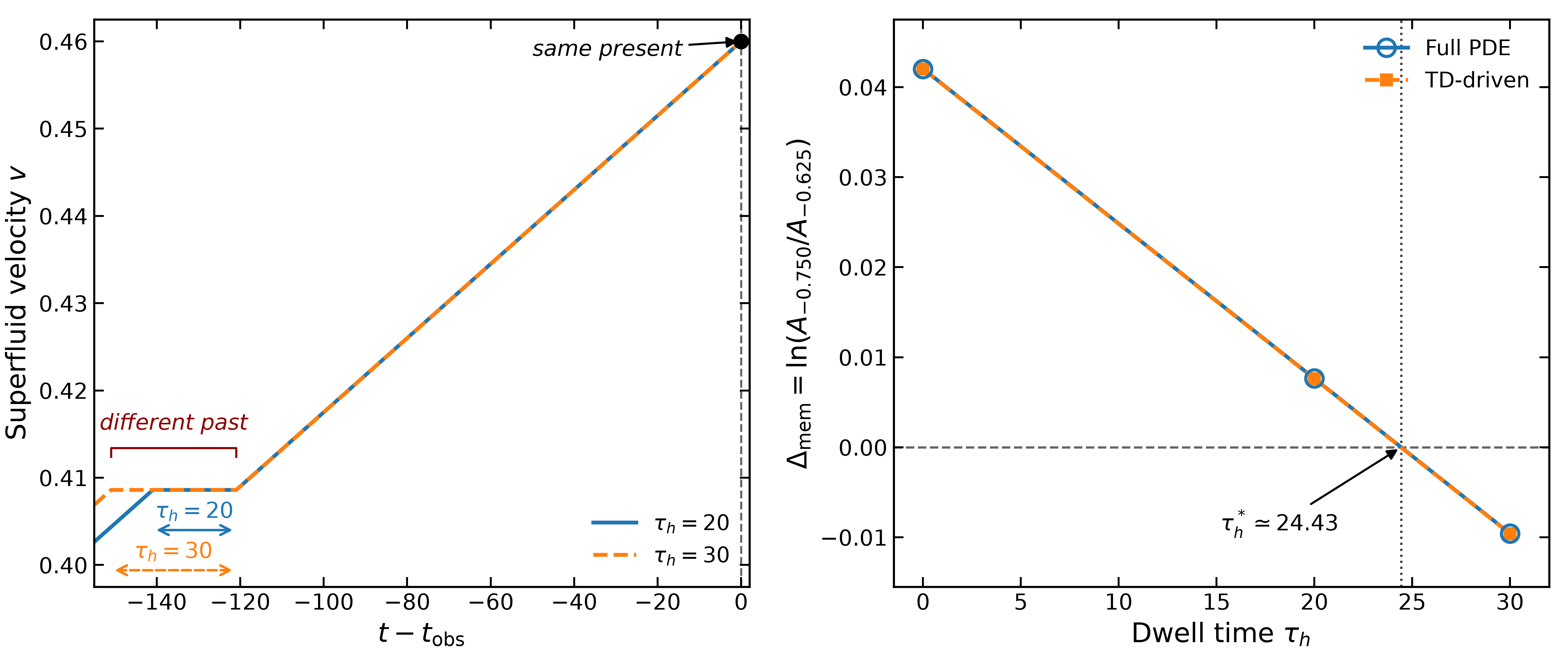}
    \caption{
        History-dependent reversal of the relative
        modal ordering.
        Left: protocols with dwell times $\tau_h=20$
        and $30$, aligned at their observation times.
        Both reach $v_{\rm obs}=0.460$ with
        $\dot v_{\rm obs}=4.25\times10^{-4}$.
        Right: the logarithmic amplitude ratio
        $\Delta_{\rm mem}
        =\ln(A_{-0.750}/A_{-0.625})$,
        including the no-dwell reference.
        Full PDE and TD-driven results nearly coincide.
        The horizontal dashed line marks equal
        amplitudes, and the vertical dotted line
        indicates the interpolated reversal at
        $\tau_h^*\simeq24.43$.
    }
    \label{fig:memory_reversal}
\end{figure}

Figure~\ref{fig:memory_reversal} shows the two
histories and the corresponding amplitude ratios.
The histories are aligned at their respective
observation times.

Linear interpolation between $\tau_h=20$ and $30$
gives
\begin{equation}
    \tau_h^{*,{\rm full}}
    =
    24.43472,
    \qquad
    \tau_h^{*,{\rm TD}}
    =
    24.41645.
    \label{eq:memory_tau_reversal}
\end{equation}

The agreement with TD-driven shows that the
reversal is captured by linear perturbation
dynamics on the evolving homogeneous background.
Nonlinear coupling between Fourier sectors is
not required in this regime.

The result concerns the ordering of two modes.
It does not establish a reversal of the
first-to-threshold winner among all allowed
momenta.

We next examine whether this history-dependent change
in relative amplitudes can be quantified from the
stationary QNM growth rates.

\section{Differential Gain and Compressed Modal Memory}
\label{sec:gain}

The reversal found in Sec.~\ref{sec:memory} shows that
the relative amplitudes of competing modes depend on
their driving history. We now examine whether this
dependence can be described by the growth accumulated
along the stationary superflow family.

\subsection{Pairwise differential gain}

For two candidate momenta $k_i$ and $k_j$, we define
the logarithmic amplitude ratio
\begin{equation}
    \Delta_{ij}(t)
    =
    \ln\frac{A_j(t)}{A_i(t)},
    \label{eq:pairwise_modal_ratio}
\end{equation}
and the difference between their stationary-QNM
growth rates,
\begin{equation}
    \Delta\gamma_{ij}(v)
    =
    \gamma_j(v)-\gamma_i(v).
    \label{eq:differential_growth}
\end{equation}

Consider two protocols, $\mathcal H$ and
$\mathcal H_0$, starting from the same initial
conditions. We compare them at the same imposed
superflow velocity and driving rate, although
their observation times may differ.

Relative to the reference protocol $\mathcal H_0$,
the excess accumulated differential gain is
\begin{equation}
    W_{ij}[\mathcal H|\mathcal H_0] = \int_{t_i}^{t_{\rm obs}[\mathcal H]} dt\, \Delta\gamma_{ij} \bigl[v_{\mathcal H}(t)\bigr] - \int_{t_i}^{t_{\rm obs}[\mathcal H_0]} dt\, \Delta\gamma_{ij} \bigl[v_{\mathcal H_0}(t)\bigr].
    \label{eq:differential_gain}
\end{equation}
The corresponding change in modal ordering is
\begin{equation}
    \delta\Delta_{ij}
    =
    \Delta_{ij}[\mathcal H]
    -
    \Delta_{ij}[\mathcal H_0],
    \label{eq:history_modal_increment}
\end{equation}
where each amplitude ratio is evaluated at
its respective observation time.

In the transported-QNM approximation of
Eq.~\eqref{eq:transported_qnm_amplitude},
the transport factors depend only on the
endpoints of the stationary background family.
They cancel between histories with common
initial and final values of $v$, leaving
\begin{equation}
    \delta\Delta_{ij}
    \simeq
    W_{ij}.
    \label{eq:memory_gain_law}
\end{equation}
This relation is exact within the specified
transported-QNM amplitude prescription.
For the full evolution it is approximate,
since the perturbations and homogeneous
background need not follow the instantaneous
stationary modes.

For the dwell protocol of
Sec.~\ref{sec:memory}, the common ramp
contributions cancel and
\begin{equation}
    W_{ij}
    =
    \Delta\gamma_{ij}(v_h)\tau_h.
    \label{eq:dwell_differential_gain}
\end{equation}
With $(k_i,k_j)=(-0.625,-0.750)$ and
$v_h=0.4086$, the stationary spectrum gives
\begin{equation}
    \Delta\gamma_{ij}(v_h)
    =
    -1.72011\times10^{-3}.
\end{equation}
Using the no-dwell amplitude ratio as the
reference, Eq.~\eqref{eq:memory_gain_law}
predicts
\begin{equation}
    \tau_h^{*,{\rm QNM}}
    =
    24.43348,
\end{equation}
compared with the interpolated full-PDE
reversal at $\tau_h^{*,{\rm full}}=24.43472$.

\subsection{Different histories at fixed differential gain}

We first vary the dwell location while
keeping the background and mode pair fixed:
\begin{equation}
    \rho=10.0417,
    \qquad
    (k_i,k_j)=(-0.625,-0.750).
\end{equation}

Three dwell locations are chosen between
the two instability thresholds,
\begin{equation}
    v_h=
    0.4073616,\,
    0.4086000,\,
    0.4095698.
\end{equation}
Their differential growth rates are
\begin{equation}
    \Delta\gamma_{ij}(v_h)
    =
    -1.84246\times10^{-3},\,
    -1.72011\times10^{-3},\,
    -1.62328\times10^{-3},
    \label{eq:three_differential_rates}
\end{equation}
respectively. At each location, the dwell
times are adjusted to give
\begin{equation}
    W_{ij}
    =
    -0.020,\,
    -0.042,\,
    -0.050,
    \label{eq:common_gains}
\end{equation}
together with the no-dwell reference.

The resulting changes in modal ordering
follow Eq.~\eqref{eq:memory_gain_law}
closely. Fits through the origin give
\begin{equation}
    \delta\Delta_{ij}^{\rm full}
    =
    0.9998808\,W_{ij},
    \qquad
    R^2=0.9999990857,
    \label{eq:full_memory_collapse}
\end{equation}
and
\begin{equation}
    \delta\Delta_{ij}^{\rm TD}
    =
    0.9998234\,W_{ij},
    \qquad
    R^2=0.9999992625.
    \label{eq:td_memory_collapse}
\end{equation}
The dependence on the dwell location is
small once the excess differential gain
is fixed.

We next compare three different histories:
a single dwell at $v=0.4086$, two dwells
at separated velocities, and a continuous
slow-ramp window. Each starts at
$v_i=0.395$ and returns to the common
rate
\begin{equation}
    \dot v=4.25\times10^{-4}.
\end{equation}
The protocols are compared at
$v_{\rm obs}=0.460$, with their parameters
chosen to produce the same $W_{ij}$.

To quantify the agreement, we define
the relative gain-law error
\begin{equation}
    \epsilon_{\rm gain}
    =
    \frac{
        |\delta\Delta_{ij}-W_{ij}|
    }{
        |W_{ij}|
    },
    \qquad W_{ij}\neq0.
    \label{eq:gain_law_error}
\end{equation}
For a group of protocols with the same
$W_{ij}$, the relative path spread is
\begin{equation}
    \epsilon_{\rm path}
    =
    \frac{
        \max_{\mathcal H}\delta\Delta_{ij}[\mathcal H]
        -
        \min_{\mathcal H}\delta\Delta_{ij}[\mathcal H]
    }{
        |W_{ij}|
    }.
    \label{eq:gain_path_spread}
\end{equation}

The largest path spread in the full
evolutions is $9.190\times10^{-4}$.
The maximum gain-law error is
$6.351\times10^{-4}$, with comparable
results from TD-driven. The errors here
measure the accuracy of the reduced
description, rather than the numerical
discretisation.

\begin{table}[t]
    \centering
    \small
    \caption{
        Differential-gain tests at
        $\rho=10.0417$ for
        $(k_i,k_j)=(-0.625,-0.750)$.
        The collapse test varies the dwell
        location. The path test compares
        a single dwell, a split dwell and
        a continuous slow-ramp window
        at fixed $W_{ij}$.
        Relative errors and spreads are
        defined in
        Eqs.~\eqref{eq:gain_law_error}
        and \eqref{eq:gain_path_spread}.
    }
    \label{tab:compressed_memory}
    \begin{tabular}{lcc}
        \hline
        Test
        & TD-driven
        & Full PDE
        \\
        \hline
        Collapse slope through origin
        & $0.9998234$
        & $0.9998808$
        \\
        Collapse $R^2$
        & $0.9999992625$
        & $0.9999990857$
        \\
        Maximum relative path spread
        & $8.902\times10^{-4}$
        & $9.190\times10^{-4}$
        \\
        Maximum relative gain-law error
        & $5.830\times10^{-4}$
        & $6.351\times10^{-4}$
        \\
        \hline
    \end{tabular}
\end{table}

The results show that these protocols
produce nearly the same change in relative
modal amplitudes when their accumulated
differential gain is matched.

\subsection{Different backgrounds and mode pairs}

We finally test the relation away from
the baseline background and mode pair.

First, we increase the fixed charge
density to
\begin{equation}
    \rho=11.5,
\end{equation}
while retaining
$(k_i,k_j)=(-0.625,-0.750)$.
The corresponding instability thresholds
are
\begin{equation}
    v_c(-0.625)=0.44867391,
    \qquad
    v_c(-0.750)=0.45080057.
\end{equation}
We use
\begin{equation}
    \dot v=1.5\times10^{-3},
    \qquad
    v_h=0.486,
    \qquad
    v_{\rm obs}=0.490.
\end{equation}
Both modes are unstable at the hold,
where
\begin{equation}
    \Delta\gamma_{ij}(v_h)
    =
    0.00276212.
\end{equation}
The differential growth rate is positive,
unlike that of the baseline dwell protocol.

The full evolution gives an interpolated
reversal time
\begin{equation}
    \tau_h^{*,{\rm full}}
    =
    34.63898,
\end{equation}
and a maximum relative gain-law error of
$1.0156\%$.

We then return to $\rho=10.0417$ and
consider the neighbouring pair
\begin{equation}
    (k_i,k_j)=(-0.750,-0.875).
\end{equation}
Its instability thresholds are
\begin{equation}
    v_c(-0.750)=0.40972035,
    \qquad
    v_c(-0.875)=0.41259167.
\end{equation}
For
\begin{equation}
    \dot v=8.0\times10^{-4},
    \qquad
    v_h=0.47359375,
    \qquad
    v_{\rm obs}=0.480,
\end{equation}
both modes are again unstable,
with
\begin{equation}
    \Delta\gamma_{ij}(v_h)
    =
    0.00203625.
\end{equation}
The full-PDE reversal occurs at
\begin{equation}
    \tau_h^{*,{\rm full}}
    =
    27.79035,
\end{equation}
and the maximum relative gain-law error
is $0.9838\%$.

Table~\ref{tab:memory_generalisation}
summarises the two tests.

\begin{table}[t]
    \centering
    \small
    \caption{
        Tests with a second fixed-density
        background and a second pair of
        competing modes.
        The reversal times are interpolated
        from the full-PDE results.
        The final column gives the maximum
        relative gain-law error over the
        tested non-zero gains.
    }
    \label{tab:memory_generalisation}
    \begin{tabular}{lcccc}
        \hline
        Case
        & $\rho$
        & $(k_i,k_j)$
        & $\tau_h^{*,{\rm full}}$
        & Maximum error
        \\
        \hline
        Second background
        & $11.5$
        & $(-0.625,-0.750)$
        & $34.639$
        & $1.0156\%$
        \\
        Second pair
        & $10.0417$
        & $(-0.750,-0.875)$
        & $27.790$
        & $0.9838\%$
        \\
        \hline
    \end{tabular}
\end{table}

The baseline-subtracted relation
\begin{equation}
    \delta\Delta_{ij}
    \simeq
    W_{ij}
    \label{eq:compressed_memory}
\end{equation}
thus remains accurate at approximately
the per-cent level in these additional
tests.

This concerns the change in pairwise
modal ordering, not the absolute
amplitudes. Matching $W_{ij}$ does not
imply identical bulk states, nor does
it determine the first-to-threshold
winner among several competing modes.
As shown in Sec.~\ref{sec:tdlinear},
absolute finite-rate amplification
depends on both radial perturbation
dynamics and the evolving homogeneous
background.

For the controlled protocols considered
here, the excess accumulated differential
QNM gain provides a useful reduced
description of the history-dependent
change in relative modal amplitudes.

\section{Conclusion and Discussion}
\label{sec:conclusion}

We have studied finite-rate Landau instability in a
probe-limit holographic superfluid, focusing on the
amplification and selection of competing
counter-propagating modes. The stationary QNM spectrum
determines the onset of each instability, but the
instantaneous growth rates alone do not determine
which mode first reaches a prescribed amplitude
during a finite-rate drive.

For eight seeded Fourier modes, adiabatic QNM transport
reproduces the full-evolution winner in eight of the
nine combinations of ramp rate and amplitude threshold
considered here. At the fastest ramp and largest
threshold, however, the full evolution selects
$k=-0.875$, whereas the transported-QNM approximation
selects $k=-0.750$. The instantaneous fastest-growing
mode is different from both. The selected mode thus
depends on the accumulated amplification and on the
dynamical corrections to stationary-QNM transport.

These corrections can be examined within linear
perturbation theory. Evolution on the actual driven
homogeneous background reproduces the full-PDE modal
amplitudes and their reordering to high accuracy.
Allowing unrestricted radial linear evolution on the
instantaneous stationary background does not give
the same result. The difference shows that the
finite-rate response of the homogeneous background
matters even when the finite-momentum perturbations
remain small.

We have investigated this response independently.
The homogeneous background departs from the stationary
family during the ramp, and its subsequent relaxation
is consistent with the slowest physical $k=0$ QNM.
A forced linear response about the moving stationary
family reproduces the background lag and, when used
in the finite-momentum tangent equations, recovers
the observed winner for the tested competing modes.
The slowest homogeneous QNM describes the scalar
response well, although other radial modes contribute
appreciably to the radial boundary response of $A_x$.

The comparison also clarifies the apparent success
of adiabatic QNM transport. Unrestricted radial
evolution on the stationary background and the
correction from the forced homogeneous response
shift the relative modal amplification in opposite
directions. Their partial cancellation helps explain
why the adiabatic approximation can be closer to
the full evolution than TD-static, despite neglecting
more of the dynamics. This cancellation is not
assumed to hold for arbitrary driving protocols.

We have also examined the effect of earlier driving
history. Protocols with identical present driving
conditions can produce opposite relative orderings
of two competing modes. For the controlled histories
studied here, the resulting change in their logarithmic
amplitude ratio is well approximated by the excess
accumulated differential QNM gain. The relation
remains accurate under the tested changes of driving
path, charge density and mode pair. It provides
a reduced description of pairwise modal memory,
rather than a general rule for absolute mode selection.

Our results concern specified initial perturbations
in the weak-amplitude regime, within the probe
limit and for longitudinal modes of homogeneous
superflows. The forced-background reconstruction
has been tested over a limited range of driving
rates and competing momenta. Its accuracy under
stronger driving, and the connection between
early-time mode selection and subsequent nonlinear
relaxation, remain separate questions. Within
the regime examined here, finite-rate mode
selection is governed not only by accumulated
QNM growth but also by the dynamical response
of the background on which the unstable modes
propagate.

\begin{acknowledgments}
The author acknowledges HIAS for access to the ``Quantum Universe
Physical Simulation Platform''. This work was supported by the
National Natural Science Foundation of China under Grant No.~12505066.
\end{acknowledgments}

\clearpage
\appendix

\section{Numerical Formulation and Validation}
\label{app:numerics}

\subsection{Time-dependent bulk equations}

We solve the matter equations in the ingoing
Eddington--Finkelstein coordinates of
Eq.~\eqref{eq:ef_metric}, using the gauge $A_z=0$
and the rescaled scalar field $\psi=\Psi/z$.
With one periodic boundary direction $x$, the
evolved variables are
\begin{equation}
    \psi(t,z,x), \qquad
    A_x(t,z,x), \qquad
    \rho(t,x).
\end{equation}
The temporal gauge field $A_t$ is reconstructed
from the radial Maxwell constraint at each
Runge--Kutta stage.

The scalar evolution equation is
\begin{equation}
    \partial_t\partial_z\psi = \frac12\partial_z \left(f\partial_z\psi\right) +\frac12\partial_x^2\psi -iA_x\partial_x\psi +iA_t\partial_z\psi -\frac{i}{2} \left(\partial_xA_x-\partial_zA_t\right)\psi -\frac12\left(z+A_x^2\right)\psi.
    \label{eq:app_scalar_evolution}
\end{equation}
The longitudinal Maxwell equation gives
\begin{equation}
    \partial_t\partial_z A_x = \frac12\partial_z \left(f\partial_zA_x\right) -|\psi|^2A_x +\operatorname{Im} \left(\psi^*\partial_x\psi\right) +\frac12\partial_x\partial_zA_t.
    \label{eq:app_ax_evolution}
\end{equation}
The radial Maxwell constraint is
\begin{equation}
    \partial_z^2A_t
    =
    \partial_z\partial_xA_x
    -
    2\operatorname{Im}
    \left(\psi^*\partial_z\psi\right).
    \label{eq:app_at_constraint}
\end{equation}

At the AdS boundary, we impose
\begin{equation}
    \psi(t,0,x)=0,\qquad A_t(t,0,x)=\mu[v(t)],\qquad A_x(t,0,x)=-v(t)\mu[v(t)],
    \label{eq:app_driven_boundary}
\end{equation}
where $\mu(v)$ is determined from the stationary
family at fixed mean charge density. During
the drive, $v(t)$ labels the imposed boundary
sources; the evolving bulk configuration need
not be stationary.

The boundary charge density is defined by
\begin{equation}
    \left.\partial_zA_t\right|_{z=0}
    =
    -\rho(t,x),
    \label{eq:app_rho_boundary}
\end{equation}
and obeys
\begin{equation}
    \partial_t\rho=-\partial_x J_x,
    \qquad
    J_x=
    \left.\partial_z A_x\right|_{z=0}
    -\partial_t a_x(t),
    \label{eq:app_charge_evolution}
\end{equation}
where $a_x(t)=A_x(t,0)$ is spatially uniform.
The second term is the boundary electric-field
contribution to the current in ingoing
Eddington--Finkelstein coordinates. Since it is
independent of $x$, it drops out of the charge
continuity equation. Periodicity preserves the
spatially averaged charge density, although
$\rho(t,x)$ may develop spatial variations.

At the future horizon $z=1$, no additional
ingoing boundary condition is imposed.
Regularity is implemented through the
Eddington--Finkelstein formulation.

\subsection{Pseudospectral discretisation}

We use a Chebyshev--Fourier pseudospectral
discretisation
\cite{Chesler:2013lia,Trefethen:2000Spectral}.
The radial direction is represented by
$N_z$ Chebyshev--Lobatto points,
\begin{equation}
    z_j
    =
    \frac12
    \left[
        1-\cos\left(
        \frac{\pi j}{N_z-1}
        \right)
    \right],
    \qquad
    j=0,\ldots,N_z-1.
    \label{eq:app_chebyshev_grid}
\end{equation}
The periodic direction uses $N_x$ Fourier points,
\begin{equation}
    x_\ell
    =
    \frac{\ell L_x}{N_x},
    \qquad
    \ell=0,\ldots,N_x-1,
    \qquad
    L_x=16\pi.
    \label{eq:app_fourier_grid}
\end{equation}
The allowed momenta are therefore
\begin{equation}
    k_n=\frac{2\pi n}{L_x}=\frac{n}{8}.
\end{equation}

Let $D$ be the Chebyshev differentiation
matrix. To integrate a radial derivative
with a prescribed boundary value, we
replace the first row of $D$ and define
\begin{equation}
    \widetilde D_{0j} =\delta_{0j},\qquad \widetilde D_{ij} =D_{ij}\qquad(i>0),\qquad \mathcal I_z =\widetilde D^{-1}.
    \label{eq:app_radial_integrator}
\end{equation}
For $\partial_z u=s$ with prescribed $u(0)$,
the first entry of $s$ is replaced by the
boundary value before applying
$\mathcal I_z$.

This operator recovers $\partial_t\psi$
and $\partial_tA_x$ from
Eqs.~\eqref{eq:app_scalar_evolution}
and \eqref{eq:app_ax_evolution}.
The temporal gauge field is reconstructed
by integrating Eq.~\eqref{eq:app_at_constraint}
twice, using
\begin{equation}
    \partial_zA_t(0,x)=-\rho(t,x),
    \qquad
    A_t(0,x)=\mu[v(t)].
\end{equation}

The integration constants in the
evolution equations are fixed by the
time derivatives of the boundary sources.
In particular,
\begin{equation}
    \left.\partial_t\psi\right|_{z=0}=0,
    \qquad
    \left.\partial_tA_x\right|_{z=0}
    =
    \frac{d}{dt}
    \left[-v(t)\mu[v(t)]\right].
    \label{eq:app_ax_source_derivative}
\end{equation}
The second term is essential during a
finite-rate ramp, since both $v$ and
$\mu(v)$ change with time.

Fourier derivatives are evaluated
spectrally, with nonlinear products
computed in physical space.
For dealiasing, Fourier modes with
integer index $|n|>N_x/3$ are removed
every four time steps. Time integration
uses the classical fourth-order
Runge--Kutta method.

The production calculations use the
baseline spatial resolution
\begin{equation}
    N_z=28,
    \qquad
    N_x=128,
    \label{eq:app_baseline_resolution}
\end{equation}
with a maximum time step
$\Delta t_{\max}=0.02$.
The actual step is also subject to
the Fourier stability bound
\begin{equation}
    \Delta t
    \leq
    \min\left(
        \Delta t_{\max},
        \frac{1}{1+k_{\rm cut}^2}
    \right),
    \label{eq:app_timestep_bound}
\end{equation}
where $k_{\rm cut}$ is the largest
retained Fourier wavenumber.
The step is adjusted so that the
integration interval is divided
into an integer number of steps.

For the eight-mode selection calculations
in Sec.~\ref{sec:selection}, the initially
seeded momenta are
\begin{equation}
    k_{-m}=-\frac{m}{8},
    \qquad m=1,\ldots,8,
\end{equation}
with equal initial boundary-condensate
amplitudes
\begin{equation}
    A_{k_{-m}}(0)=10^{-5}.
\end{equation}
The numerical dependence of the mode
selection on the radial resolution,
Fourier resolution and time step is
examined in Appendix~\ref{app:robustness}.

\subsection{QNM--time-domain validation}

We validate the time evolution against
an independently computed QNM on a fixed
background. The benchmark uses
\begin{equation}
    \rho=10.0417,
    \qquad
    v=0.415,
    \qquad
    k=-0.500,
    \label{eq:app_validation_point}
\end{equation}
with an initial QNM perturbation
of amplitude $\epsilon=10^{-5}$.

The frequency-domain eigenvalue is
\begin{equation}
    \omega_{\rm QNM}
    =
    -0.0200642664022
    +
    0.00526610527994\,i.
    \label{eq:app_qnm_frequency}
\end{equation}
We evolve the full nonlinear bulk equations
at fixed $v$ and fit the complex Fourier
coefficient of the boundary condensate to
\begin{equation}
    \mathcal O_k(t)
    \propto
    e^{-i\omega_{\rm evo}t},
    \label{eq:app_time_fit}
\end{equation}
over $20\leq t\leq280$. The fitted
frequency is
\begin{equation}
    \omega_{\rm evo}
    =
    -0.0200642664810
    +
    0.00526610531779\,i.
    \label{eq:app_evolution_frequency}
\end{equation}

The relative differences in the real
and imaginary parts are
$3.9\times10^{-9}$ and
$7.2\times10^{-9}$, respectively.
Table~\ref{tab:qnm_time_validation}
also lists the charge-conservation,
constraint and boundary diagnostics
for this benchmark.

\begin{table}[t]
    \centering
    \caption{
        Single-mode QNM--time-domain
        validation at $\rho=10.0417$,
        $v=0.415$ and $k=-0.500$.
        The evolution uses $N_z=28$,
        $N_x=128$, $\Delta t_{\max}=0.02$
        and an initial QNM amplitude
        of $10^{-5}$.
    }
    \label{tab:qnm_time_validation}
    \begin{tabular}{lc}
        \hline
        Diagnostic & Value \\
        \hline
        $\operatorname{Re}\omega_{\rm QNM}$
        & $-0.0200642664022$ \\
        $\operatorname{Re}\omega_{\rm evo}$
        & $-0.0200642664810$ \\
        Relative error in
        $\operatorname{Re}\omega$
        & $3.9\times10^{-9}$ \\
        $\operatorname{Im}\omega_{\rm QNM}$
        & $0.00526610527994$ \\
        $\operatorname{Im}\omega_{\rm evo}$
        & $0.00526610531779$ \\
        Relative error in
        $\operatorname{Im}\omega$
        & $7.2\times10^{-9}$ \\
        Maximum relative charge drift
        & $5.3\times10^{-16}$ \\
        Maximum Maxwell-constraint residual
        & $5.0\times10^{-12}$ \\
        Maximum boundary-condition error
        & $1.6\times10^{-14}$ \\
        QNM eigenproblem residual
        & $\sim10^{-16}$ \\
        \hline
    \end{tabular}
\end{table}

The agreement tests the stationary
background, QNM eigenproblem,
radial constraint reconstruction
and time-domain evolution within
a common numerical formulation.

This single-mode benchmark is distinct
from the convergence tests of the
eight-mode competition. The latter
require the relative ordering of closely
competing threshold crossings to remain
stable under numerical refinement.
Those tests, together with further
constraint and robustness checks,
are presented in
Appendix~\ref{app:robustness}.

\section{Finite-Rate QNM Transport}
\label{app:qnm_transport}

The transported-QNM approximation used in
Sec.~\ref{sec:selection} requires a consistent
normalisation of the instantaneous QNM eigenvectors
and their projection onto the boundary condensate.
We describe the construction and its numerical
validation below.

\subsection{Biorthogonal parallel transport}

For fixed momentum $k$ and superflow parameter $v$,
the discretised QNM problem takes the form
\begin{equation}
    \mathsf A_k(v)r_k(v)
    +\omega_k(v)\mathsf B_k(v)r_k(v)=0,
    \label{eq:app_qnm_pencil}
\end{equation}
with left eigenvector satisfying
\begin{equation}
    l_k^\dagger(v)\mathsf A_k(v)
    +\omega_k(v)l_k^\dagger(v)\mathsf B_k(v)=0.
\end{equation}
We impose the biorthogonal normalisation
\begin{equation}
    l_k^\dagger(v)\mathsf B_k(v)r_k(v)=1.
    \label{eq:app_biorthogonal_norm}
\end{equation}

At the first point of each QNM family, we select
the counter-propagating sound branch. At subsequent
points, the branch is followed by choosing the
frequency nearest to the previous one within
the sound-mode pair.

The left and right eigenvectors retain a complex
normalisation freedom even after imposing
Eq.~\eqref{eq:app_biorthogonal_norm}. We fix their
relative phase along the QNM family using a
discrete biorthogonal parallel-transport prescription.

On a grid $v^{(0)}<v^{(1)}<\cdots$, the left and
right eigenvectors at each new point are rotated
by the same unit phase. This preserves their
biorthogonal normalisation. The phase is chosen
so that the overlap
\begin{equation}
    S_{k,m}\equiv
    l_{k,m}^{\dagger}\mathsf B_{k,m}r_{k,m-1}>0,
    \qquad m\geq1,
    \label{eq:app_pt_overlap}
\end{equation}
is real and positive. This choice is well defined
provided the overlap is non-zero along the
tracked branch.

The accumulated modal transport is
\begin{equation}
    \mathcal T_{k,m}=\prod_{n=1}^{m}S_{k,n},
    \qquad \mathcal T_{k,0}=1.
    \label{eq:app_modal_transport}
\end{equation}

In the smooth limit, this construction corresponds
to the single-mode adiabatic evolution
\begin{equation}
    \frac{d}{dv}\ln c_k
    =-l_k^\dagger\mathsf B_k\partial_v r_k,
    \label{eq:app_transport_connection}
\end{equation}
where $c_k$ is the instantaneous modal
coefficient. Transitions to other radial QNMs
are neglected.

The observable used for mode selection is the
one-sided Fourier coefficient of the boundary
condensate. If $q_R$ and $q_I$ are the scalar
components of $r_k$, its boundary weight is
\begin{equation}
    C_{k,m}=\frac12
    \left.\partial_z(q_R+iq_I)\right|_{z=0}.
    \label{eq:app_observable_weight}
\end{equation}
The factor $1/2$ follows from the one-sided
Fourier convention used to construct the
real perturbation in the time evolution.

The corresponding transport factor is
\begin{equation}
    F^{\rm PT}_{k,m}
    =\frac{C_{k,m}}{C_{k,0}}\,
    \mathcal T_{k,m}.
    \label{eq:app_complex_transport}
\end{equation}

To see that this factor is independent of the
eigenvector normalisation, consider an arbitrary
non-zero complex rescaling at each grid point,
\begin{equation}
    r_{k,m}\mapsto a_m r_{k,m},
    \qquad
    l_{k,m}^{\dagger}\mapsto
    a_m^{-1}l_{k,m}^{\dagger}.
\end{equation}
This preserves Eq.~\eqref{eq:app_biorthogonal_norm}.
The overlap and boundary weight transform as
\begin{equation}
    S_{k,m}\mapsto
    \frac{a_{m-1}}{a_m}S_{k,m},
    \qquad
    C_{k,m}\mapsto a_m C_{k,m}.
\end{equation}
The rescaling factors telescope in
$\mathcal T_{k,m}$ and cancel those of the
boundary-weight ratio. Hence
$F^{\rm PT}_{k,m}$ is invariant under the
local complex normalisation freedom.

For the amplitude calculations, we retain
\begin{equation}
    \mathcal P_k\!\left(v^{(m)}\right)
    =\left|F^{\rm PT}_{k,m}\right|,
    \qquad
    \mathcal P_k\!\left(v^{(0)}\right)=1.
    \label{eq:app_amplitude_transport}
\end{equation}
Here $v^{(0)}$ is the reference point of the
QNM table. The ratio
$\mathcal P_k(v)/\mathcal P_k(v_i)$
is independent of this choice.

Between tabulated velocities, the QNM growth
rate $\gamma_k(v)$ and $\ln\mathcal P_k(v)$
are evaluated using piecewise cubic Hermite
interpolation.

For a prescribed history $v(t)$, the
transported amplitude is
\begin{equation}
    \ln\frac{A_k^{\rm tr}(t)}{A_k(t_i)}
    =\int_{t_i}^{t}\gamma_k[v(t')]\,dt'
    +\ln\frac{\mathcal P_k[v(t)]}
             {\mathcal P_k[v(t_i)]}.
    \label{eq:app_transport_general}
\end{equation}
For a linear ramp $\dot v=R$, this reduces
to Eq.~\eqref{eq:transported_qnm_amplitude}.

The corresponding instantaneous logarithmic
growth rate is
\begin{equation}
    g_k^{\rm PT}(t)
    \equiv\frac{d}{dt}\ln A_k^{\rm tr}(t)
    =\gamma_k[v(t)]
    +\dot v(t)\,\partial_v\ln\mathcal P_k(v).
    \label{eq:app_local_growth}
\end{equation}

Only the modulus of the transport factor
enters the amplitude predictions considered
in this paper. The discrete construction
also produces a complex transport phase,
but the amplitude validation below does
not establish the accuracy of that phase
in a driven evolution.

\subsection{Single-mode validation}

We test the transport prescription against
an independent single-mode finite-rate
evolution. The calculation uses
\begin{equation}
    \rho=10.0417,
    \qquad
    k=-0.500,
    \qquad
    v:0.385\rightarrow0.425,
    \label{eq:app_transport_test_setup}
\end{equation}
with an initial boundary-condensate
amplitude of $10^{-5}$.

The instability threshold for this mode is
\begin{equation}
    v_c(k=-0.500)=0.4051007.
    \label{eq:app_transport_vc}
\end{equation}

The test uses four ramp rates,
\begin{equation}
    R=
    10^{-4},\,
    2\times10^{-4},\,
    5\times10^{-4},\,
    10^{-3}.
    \label{eq:app_transport_test_rates}
\end{equation}
These rates belong to the single-mode
validation and are distinct from the
three rates used in the eight-mode
selection analysis.

The stationary QNM family is sampled at
81 velocities. Between the endpoints,
the boundary-observable weight changes
little, whereas the accumulated modal
transport produces a larger correction:
\begin{equation}
    \left|
    \frac{C_k(v_f)}{C_k(v_i)}
    \right|
    =
    1.00078,
    \qquad
    \left|
    \mathcal T_k(v_f)
    \right|
    =
    0.92079.
\end{equation}
Their product gives
\begin{equation}
    \frac{\mathcal P_k(v_f)}
         {\mathcal P_k(v_i)}
    =
    0.92151.
    \label{eq:app_transport_endpoint}
\end{equation}

We compare the direct bulk evolution with
two approximations. The growth-only
prediction retains the integrated QNM
growth rate but sets the amplitude
transport factor to unity. The transported
prediction includes both terms in
Eq.~\eqref{eq:app_transport_general}.

For either prediction, the relative
amplitude error is defined by
\begin{equation}
    \epsilon_A
    =
    \left|
    \frac{A_k^{\rm direct}}
         {A_k^{\rm pred}}
    -1
    \right|.
    \label{eq:app_transport_amp_error}
\end{equation}
The final errors are evaluated after
the prescribed ramp and subsequent
hold at $v_f$.

We also compare the local logarithmic
growth rates. The direct rate is obtained
from a local polynomial fit to
$\ln A_k(t)$, while the two predictions
use $\gamma_k(v)$ and
Eq.~\eqref{eq:app_local_growth},
respectively. The RMS errors are
evaluated over the interior of the ramp.

\begin{table}[t]
    \centering
    \small
    \caption{
        Single-mode validation of finite-rate
        QNM transport for $k=-0.500$ and
        $v:0.385\rightarrow0.425$.
        The amplitude columns give the final
        relative error with respect to direct
        bulk evolution. The last two columns
        give the RMS error of the local
        logarithmic growth rate during the ramp.
    }
    \label{tab:qnm_transport_validation}
    \begin{tabular}{ccccc}
        \hline
        $R$
        & \shortstack{Growth-only\\amplitude error}
        & \shortstack{Transported\\amplitude error}
        & \shortstack{Growth-only\\growth-rate RMS}
        & \shortstack{Transported\\growth-rate RMS}
        \\
        \hline
        $1\times10^{-4}$
        & $7.013\%$
        & $0.908\%$
        & $1.905\times10^{-4}$
        & $4.757\times10^{-5}$
        \\
        $2\times10^{-4}$
        & $7.012\%$
        & $0.909\%$
        & $3.785\times10^{-4}$
        & $9.246\times10^{-5}$
        \\
        $5\times10^{-4}$
        & $7.005\%$
        & $0.917\%$
        & $9.254\times10^{-4}$
        & $2.380\times10^{-4}$
        \\
        $1\times10^{-3}$
        & $6.976\%$
        & $0.947\%$
        & $1.785\times10^{-3}$
        & $4.910\times10^{-4}$
        \\
        \hline
    \end{tabular}
\end{table}

Table~\ref{tab:qnm_transport_validation}
shows that including the transport factor
reduces the final amplitude error from
approximately $7\%$ to below $1\%$
for all four rates. The local growth-rate
RMS error is reduced by a factor of
approximately $3.6$--$4.1$.

The eigenvector transport is also
numerically well resolved. Along the
QNM family,
\begin{equation}
    \max_v
    \left|
        l_k^\dagger\mathsf B_k r_k-1
    \right|
    =
    5.70\times10^{-15},
\end{equation}
and
\begin{equation}
    \max_m
    \left|
        \operatorname{Im}S_{k,m}
    \right|
    =
    4.86\times10^{-16}.
\end{equation}
The largest right-QNM residual is
$2.27\times10^{-16}$.

The test supports the amplitude transport
prescription used in Sec.~\ref{sec:selection}.
It accounts for the changing eigenvector
normalisation and boundary projection
along the stationary family.

It does not, however, include transitions
between different radial QNMs or the
finite-rate departure of the homogeneous
background from that family. The effects
of these approximations are examined
through time-dependent linear evolution
in Sec.~\ref{sec:tdlinear}.

\section{Time-Dependent Linearisation and Forced Backgrounds}
\label{app:td_linear}

We give the linearised equations used in
Sec.~\ref{sec:tdlinear}, including the forced
homogeneous response and its phase fixing.
We also describe the physical $k=0$
relaxation spectrum and the instantaneous-QNM
decomposition of finite-momentum perturbations.

\subsection{Linearised evolution in a fixed momentum sector}

Let the homogeneous background be
\begin{equation}
    \psi_{\rm b}(t,z)
    =
    R_{\rm b}(t,z)+iI_{\rm b}(t,z),
    \qquad
    A_{t{\rm b}}(t,z),
    \qquad
    A_{x{\rm b}}(t,z).
    \label{eq:app_td_background}
\end{equation}

For a perturbation proportional to $e^{ikx}$,
we define
\begin{equation}
    \delta P=\delta R+i\delta I,
    \qquad
    \delta Q=\delta R-i\delta I.
    \label{eq:app_delta_pq}
\end{equation}
Here $\delta R$ and $\delta I$ are complex
Fourier coefficients, so $\delta Q$ need not
equal $\delta P^*$.

The linearised Maxwell constraint is
\begin{equation}
    \partial_z^2\delta A_t = ik\,\partial_z\delta A_x - 2\left( \delta R\,\partial_z I_{\rm b} +R_{\rm b}\,\partial_z\delta I \right. \left.-\delta I\,\partial_z R_{\rm b} -I_{\rm b}\,\partial_z\delta R \right),
    \label{eq:app_delta_at_constraint}
\end{equation}
with
\begin{equation}
    \delta A_t(0)=0,
    \qquad
    \partial_z\delta A_t(0)=-\delta\rho.
    \label{eq:app_delta_at_boundary}
\end{equation}
Thus $\delta A_t$ is reconstructed from
the independently evolved variables
$(\delta R,\delta I,\delta A_x,\delta\rho)$.

Define
\begin{equation}
    \mathcal D_{\pm}^{(k)}\chi = \frac12\partial_z \left(f\partial_z\chi\right) -\frac{k^2}{2}\chi \pm kA_{x{\rm b}}\chi \pm iA_{t{\rm b}}\partial_z\chi \pm\frac{i}{2} \left(\partial_zA_{t{\rm b}}\right)\chi -\frac12 \left(z+A_{x{\rm b}}^2\right)\chi.
    \label{eq:app_td_operator}
\end{equation}
The scalar tangent equations are
\begin{equation}
    \partial_t\partial_z\delta P = \mathcal D_+^{(k)}\delta P +i\delta A_t\partial_z\psi_{\rm b} + \left( \frac{k}{2}-A_{x{\rm b}} \right) \delta A_x\psi_{\rm b} + \frac{i}{2} \left(\partial_z\delta A_t\right) \psi_{\rm b},
    \label{eq:app_tangent_p}
\end{equation}
and
\begin{equation}
    \partial_t\partial_z\delta Q = \mathcal D_-^{(k)}\delta Q -i\delta A_t\partial_z\psi_{\rm b}^* - \left( \frac{k}{2}+A_{x{\rm b}} \right) \delta A_x\psi_{\rm b}^* - \frac{i}{2} \left(\partial_z\delta A_t\right) \psi_{\rm b}^*.
    \label{eq:app_tangent_q}
\end{equation}

The spatial gauge-field perturbation satisfies
\begin{equation}
    \partial_t\partial_z\delta A_x = \frac12\partial_z \left(f\partial_z\delta A_x\right) -|\psi_{\rm b}|^2\delta A_x - 2A_{x{\rm b}} \left( R_{\rm b}\delta R +I_{\rm b}\delta I \right) + ik\left( R_{\rm b}\delta I -I_{\rm b}\delta R \right) + \frac{ik}{2} \partial_z\delta A_t.
    \label{eq:app_tangent_ax}
\end{equation}
Boundary charge conservation gives
\begin{equation}
    \partial_t\delta\rho
    =
    -ik
    \left.
    \partial_z\delta A_x
    \right|_{z=0}.
    \label{eq:app_tangent_charge}
\end{equation}

All finite-momentum perturbations are
source-free:
\begin{equation}
    \delta P(0)=\delta Q(0)
    =\delta A_t(0)=\delta A_x(0)=0.
    \label{eq:app_tangent_sources}
\end{equation}
Radial integration and Maxwell-constraint
reconstruction use the operators described
in Appendix~\ref{app:numerics}.

Each momentum sector is initialised with
its tracked Landau QNM at $v_i=0.395$,
normalised to
\begin{equation}
    A_k(0)=10^{-5}.
    \label{eq:app_tangent_seed}
\end{equation}
The full real-space perturbation is constructed
as the real part of the complex QNM profile.
Consequently, its one-sided Fourier coefficient
contains a factor of $1/2$.

The original tangent-system validation uses
six momenta,
\begin{equation}
    k=-\frac{m}{8},
    \qquad m=3,\ldots,8,
\end{equation}
at $N_z=28$ and $\Delta t=0.02$.
These are a subset of the eight momenta
used in Sec.~\ref{sec:selection}.

On the initial stationary background, the
integrated tangent operator must reproduce
the QNM frequency:
\begin{equation}
    \mathcal F_k[\delta Y_k]
    =
    -i\omega_k\delta Y_k,
    \qquad
    \delta Y_k=
    (\delta R,\delta I,\delta A_x,\delta\rho).
    \label{eq:app_tangent_qnm_check}
\end{equation}
Across the six tested sectors, the maximum
normalised residual is
\begin{equation}
    \max_k\epsilon_k^{\rm QNM}
    =
    1.70\times10^{-10}.
    \label{eq:app_tangent_qnm_residual}
\end{equation}
The maximum difference between the reconstructed
$\delta A_t$ and the frequency-domain QNM
eigenvector is
\begin{equation}
    3.21\times10^{-17}.
    \label{eq:app_tangent_constraint_mismatch}
\end{equation}
These checks validate the tangent equations
and constraint reconstruction.

\subsection{TD-static and TD-driven backgrounds}

The TD-static and TD-driven calculations use
the same tangent equations. They differ
only in the homogeneous background.

For TD-static,
\begin{equation}
    X_{\rm b}(t)
    =
    X_0[v(t)],
    \label{eq:app_td_static_background}
\end{equation}
where $X_0(v)$ is the stationary family
at fixed charge density.

For TD-driven, the background is evolved
independently under the same boundary sources
as the full PDE. Writing
\begin{equation}
    X_{\rm d}(t)
    =
    \left(
        \psi_{\rm d},
        A_t^{\rm d},
        A_x^{\rm d}
    \right),
\end{equation}
the homogeneous equations are
\begin{equation}
    \partial_t\partial_z\psi_{\rm d} = \frac12\partial_z \left(f\partial_z\psi_{\rm d}\right) +iA_t^{\rm d}\partial_z\psi_{\rm d} + \frac{i}{2} \left(\partial_zA_t^{\rm d}\right) \psi_{\rm d} - \frac12 \left(z+(A_x^{\rm d})^2\right) \psi_{\rm d},
    \label{eq:app_td_driven_scalar}
\end{equation}
\begin{equation}
    \partial_t\partial_zA_x^{\rm d}
    =
    \frac12\partial_z
    \left(f\partial_zA_x^{\rm d}\right)
    -
    |\psi_{\rm d}|^2A_x^{\rm d},
    \label{eq:app_td_driven_ax}
\end{equation}
and
\begin{equation}
    \partial_z^2A_t^{\rm d}
    =
    -2\operatorname{Im}
    \left(
        \psi_{\rm d}^*
        \partial_z\psi_{\rm d}
    \right).
    \label{eq:app_td_driven_at}
\end{equation}

The boundary conditions are
\begin{equation}
    \psi_{\rm d}(t,0)=0,\qquad A_t^{\rm d}(t,0)=\mu[v(t)],\qquad \partial_zA_t^{\rm d}(t,0)=-\rho,\qquad A_x^{\rm d}(t,0)=-v(t)\mu[v(t)].
    \label{eq:app_td_driven_boundary}
\end{equation}
For the baseline calculation,
$\rho=10.0417$.

The time derivative of the $A_x$
boundary source enters the radial
integration of its evolution equation:
\begin{equation}
    \left.
    \partial_tA_x^{\rm d}
    \right|_{z=0}
    =
    \dot v\,
    \frac{d}{dv}
    \left[-v\mu(v)\right].
    \label{eq:app_driven_source_derivative}
\end{equation}
This term is retained at every
Runge--Kutta stage.

Substituting $X_0[v(t)]$ or $X_{\rm d}(t)$
into the tangent equations gives
TD-static or TD-driven, respectively.
Since both backgrounds are homogeneous,
different Fourier sectors remain
decoupled at linear order.

\subsection{Forced homogeneous response and phase fixing}

We construct the homogeneous response
about the moving stationary family.

Let
\begin{equation}
    Y=(\psi,A_x)
\end{equation}
denote the independent homogeneous fields.
The temporal gauge field is reconstructed
from the radial Maxwell constraint at
fixed charge density.

After radial integration, the evolution
can be written as
\begin{equation}
    \partial_tY
    =
    \mathcal F[Y;v,\dot v].
    \label{eq:app_homogeneous_flow}
\end{equation}
For a linear ramp, we write
$\mathcal F_R=\mathcal F[Y;v,R]$.

The stationary family satisfies
\begin{equation}
    \mathcal F_0[Y_0(v);v]=0,
    \qquad
    Y_0(v)=
    \left(
        \psi_0(v),A_{x,0}(v)
    \right).
    \label{eq:app_stationary_flow}
\end{equation}

We write the driven fields as
\begin{equation}
    \psi_{\rm F}(t,z) = e^{i\Theta(t)} \left[ \psi_0(v(t),z)+\delta\psi(t,z) \right],\qquad A_{x,\rm F}(t,z) = A_{x,0}(v(t),z) +\delta A_x(t,z).
    \label{eq:app_forced_background_expansion}
\end{equation}
The scalar phase $\Theta(t)$ is
separated from the response
$\delta Y=(\delta\psi,\delta A_x)$.

The homogeneous evolution is real-linear
in its perturbations. We therefore define
the Jacobian on the real state variables
$(\operatorname{Re}\delta\psi,
\operatorname{Im}\delta\psi,\delta A_x)$:
\begin{equation}
    \mathcal J_0(v)
    =
    D_Y\mathcal F_0[Y_0(v);v].
    \label{eq:app_homogeneous_jacobian}
\end{equation}
The derivative is taken at fixed boundary
sources and fixed charge density.

To first order in the departure from
the stationary family,
\begin{equation}
    \partial_t\delta Y = \mathcal J_0(v)\delta Y + \mathcal F_R[Y_0(v);v] - R\,\partial_vY_0(v) - \dot\Theta\,g_0(v),
    \label{eq:app_forced_homogeneous_response}
\end{equation}
where
\begin{equation}
    g_0(v)
    =
    \begin{pmatrix}
        i\psi_0(v)\\
        0
    \end{pmatrix}
    \label{eq:app_phase_generator}
\end{equation}
generates a global scalar-phase
rotation. Terms quadratic in the
response are neglected.

The forcing is
\begin{equation}
    f_R(v)
    =
    \mathcal F_R[Y_0(v);v]
    -
    R\,\partial_vY_0(v).
    \label{eq:app_forced_source}
\end{equation}
It contains both the explicit
rate-dependent boundary forcing
and the motion of the stationary family.

To make the first contribution explicit,
write the spatial gauge-field source as
\begin{equation}
    a_x(v)=-v\mu(v).
\end{equation}
In the radial-integration scheme,
$\partial_tA_x(0)=Ra_x'(v)$.
At an exact stationary profile, the
bulk right-hand sides vanish, leaving
\begin{equation}
    \mathcal F_R[Y_0(v);v]
    =
    R
    \begin{pmatrix}
        0\\
        a_x'(v)\,\mathbf 1
    \end{pmatrix},
    \label{eq:app_explicit_rate_forcing}
\end{equation}
where $\mathbf 1$ denotes the constant
radial profile.

Consequently,
\begin{equation}
    f_R(v)
    =
    R
    \begin{pmatrix}
        -\partial_v\psi_0(v)\\
        a_x'(v)\,\mathbf 1
        -\partial_vA_{x,0}(v)
    \end{pmatrix}.
    \label{eq:app_forced_source_explicit}
\end{equation}
The response has homogeneous
boundary conditions,
\begin{equation}
    \delta\psi(0)=0,
    \qquad
    \delta A_x(0)=0.
    \label{eq:app_forced_zero_sources}
\end{equation}

The temporal gauge-field response
is reconstructed from
\begin{equation}
    \partial_z^2\delta A_t = -2\operatorname{Im} \left( \delta\psi^* \partial_z\psi_0 + \psi_0^* \partial_z\delta\psi \right),
    \label{eq:app_forced_at_constraint}
\end{equation}
with
\begin{equation}
    \delta A_t(0)=0,
    \qquad
    \partial_z\delta A_t(0)=0.
    \label{eq:app_forced_at_boundary}
\end{equation}

The Jacobian has a zero mode
corresponding to a constant global
scalar-phase rotation. We remove
this freedom by imposing
\begin{equation}
    p_v(\delta Y)=0,
    \label{eq:app_forced_phase_condition}
\end{equation}
where
\begin{equation}
    p_v(\delta Y)
    =
    \operatorname{Im}
    \int_0^1 dz\,
    \psi_0^*(v,z)\delta\psi(z).
    \label{eq:app_phase_functional}
\end{equation}
This is a real-linear functional
on the homogeneous state space.

Define
\begin{equation}
    \mathcal N_\psi(v)
    =
    p_v(g_0)
    =
    \int_0^1 dz\,|\psi_0(v,z)|^2.
    \label{eq:app_phase_norm}
\end{equation}
Differentiating the phase condition gives
\begin{equation}
    p_v(\partial_t\delta Y)
    +
    R\,(\partial_vp_v)(\delta Y)=0.
    \label{eq:app_phase_condition_derivative}
\end{equation}
Substituting
Eq.~\eqref{eq:app_forced_homogeneous_response},
we obtain
\begin{equation}
    \dot\Theta
    =
    \frac{
        p_v(\mathcal J_0\delta Y+f_R)
        +
        R\,(\partial_vp_v)(\delta Y)
    }{
        \mathcal N_\psi
    }.
    \label{eq:app_forced_phase_velocity}
\end{equation}
Here
\begin{equation}
    (\partial_vp_v)(\delta Y)
    =
    \operatorname{Im}
    \int_0^1 dz\,
    (\partial_v\psi_0^*)\delta\psi.
    \label{eq:app_phase_functional_derivative}
\end{equation}

These equations determine the
phase-fixed response and the
accumulated phase $\Theta(t)$.
The forced background is then
reconstructed using
Eq.~\eqref{eq:app_forced_background_expansion}
and the Maxwell constraint.

The reconstruction retains the
complete homogeneous radial
response at linear order. It does
not assume that a single
$k=0$ QNM dominates.

Finally, the forced background
is used in the finite-momentum
tangent system,
\begin{equation}
    \partial_t\delta X_k
    =
    \mathcal L_k[X_{\rm F}(t)]
    \delta X_k.
    \label{eq:app_forced_finite_k}
\end{equation}
The finite-momentum equations are
the same as in TD-static and
TD-driven; only the background differs.

\subsection{Physical \texorpdfstring{$k=0$}{k=0} modes and dynamical projection}

The physical homogeneous relaxation
spectrum is obtained after removing
the global phase zero mode.

It is useful to introduce the
phase-fixing projector
\begin{equation}
    \mathcal P_v u
    =
    u-
    g_0(v)
    \frac{p_v(u)}{\mathcal N_\psi(v)}.
    \label{eq:app_phase_projector}
\end{equation}
It satisfies
\begin{equation}
    \mathcal P_v^2=\mathcal P_v,
    \qquad
    \mathcal P_vg_0=0,
    \qquad
    p_v(\mathcal P_vu)=0.
\end{equation}

At fixed $v$, the physical
homogeneous tangent operator is
the restriction
\begin{equation}
    \mathcal J_\perp(v)
    =
    \left.
    \mathcal P_v\mathcal J_0(v)
    \right|_{\ker p_v}.
    \label{eq:app_projected_k0_jacobian}
\end{equation}
The physical eigenvalues can
equivalently be obtained from
the full Jacobian after identifying
and excluding the global phase
zero mode.

Since the state variables are
real, the Jacobian and projector
are real-linear operators.
They are complexified only when
computing eigenvalues and
biorthogonal eigenvectors.

The eigenvalue problem is
\begin{equation}
    \mathcal J_\perp(v)r_n(v)
    =
    \lambda_n(v)r_n(v),
    \qquad
    \lambda_n=-i\omega_n.
    \label{eq:app_k0_eigenproblem}
\end{equation}
For the least damped physical mode,
denoted by $s$, the relaxation
time is
\begin{equation}
    \tau_0(v)
    =
    -\frac{1}{
        \operatorname{Re}\lambda_s(v)
    }.
    \label{eq:app_k0_relaxation}
\end{equation}

During the ramp, the projector
itself depends on $v(t)$.
Using the phase condition and
Eq.~\eqref{eq:app_forced_phase_velocity},
the forced response satisfies
\begin{equation}
    \partial_t\delta Y = \mathcal P_v \left( \mathcal J_0\delta Y+f_R \right) - \frac{R}{\mathcal N_\psi} g_0\,(\partial_vp_v)(\delta Y).
    \label{eq:app_projected_forced_response}
\end{equation}
The last term ensures that the
phase condition remains satisfied
as the stationary background changes.

At fixed $v$, choose left and
right physical eigenvectors
such that
\begin{equation}
    \mathcal J_\perp r_n =\lambda_n r_n,\qquad l_n^\dagger\mathcal J_\perp =\lambda_n l_n^\dagger,\qquad l_n^\dagger r_n=1.
    \label{eq:app_k0_biorthogonal}
\end{equation}
The left eigenvectors are understood
as dual vectors on the phase-fixed
state space. They may be extended
to the full state space with
\begin{equation}
    l_n^\dagger g_0=0.
    \label{eq:app_k0_left_phase}
\end{equation}

The instantaneous coefficient
of the slow mode is
\begin{equation}
    c_s(t)
    =
    l_s^\dagger[v(t)]\,\delta Y(t),
    \label{eq:app_k0_projection}
\end{equation}
with reconstructed contribution
\begin{equation}
    \delta Y_s(t)
    =
    c_s(t)\,r_s[v(t)].
    \label{eq:app_k0_slow_reconstruction}
\end{equation}

Differentiating $c_s$ and using
Eq.~\eqref{eq:app_projected_forced_response}
gives
\begin{equation}
    \dot c_s = \lambda_s c_s + l_s^\dagger\mathcal P_v f_R + \dot l_s^\dagger\delta Y.
    \label{eq:app_k0_modal_coefficient}
\end{equation}
The term proportional to
$(\partial_vp_v)(\delta Y)$
drops out because
$l_s^\dagger g_0=0$.

The last term accounts for the
changing eigenbasis. Its numerical
value depends on the normalisation
of the instantaneous eigenvectors
and is not independently observable.

We assess the slow-mode
reconstruction using the scalar
modulus. To first order,
\begin{equation}
    \delta|\psi|
    =
    \operatorname{Re}
    \left(
        \frac{\psi_0^*}{|\psi_0|}
        \delta\psi
    \right),
    \label{eq:app_k0_scalar_modulus}
\end{equation}
where the expression is evaluated
away from zeros of $\psi_0$.

Let $\delta|\psi|_s$ be the
contribution from $\delta Y_s$.
We define
\begin{equation}
    \epsilon_{\psi,s}
    =
    \frac{
        \left\|
        \delta|\psi|-\delta|\psi|_s
        \right\|_{L^2_z}
    }{
        \left\|
        \delta|\psi|
        \right\|_{L^2_z}
    }.
    \label{eq:app_k0_scalar_residual}
\end{equation}

For the spatial gauge-field response, we use
the radial boundary coefficients,
\begin{equation}
    \delta j_x^{\rm rad}
    =
    \left.\partial_z\delta A_x\right|_{z=0},
    \qquad
    \delta j_{x,s}^{\rm rad}
    =
    \left.\partial_z\delta A_{x,s}\right|_{z=0}.
    \label{eq:app_k0_current_projection}
\end{equation}
The signed ratio
$\delta j_{x,s}^{\rm rad}/\delta j_x^{\rm rad}$
may exceed unity when other radial modes
contribute with the opposite sign. This
projection concerns the radial response
coefficient. The physical boundary-current
difference from an instantaneous stationary
state also contains the imposed term
$-\partial_t a_x(t)$, which is not part
of this modal projection.

These projections diagnose the
homogeneous response. The
finite-momentum calculations do
not use a truncated $k=0$
mode expansion.

\subsection{Instantaneous-QNM decomposition of finite-momentum modes}

We also examine the radial content
of selected TD-driven finite-momentum
perturbations.

Let
\begin{equation}
    q_k(t)
    =
    \left(
        \delta R,\delta I,
        \delta A_t,\delta A_x
    \right)_k
    \label{eq:app_projection_state}
\end{equation}
be the tangent state, including
the reconstructed $\delta A_t$.

At the instantaneous stationary
background $X_0[v(t)]$, the
generalised QNM problem gives
right and left eigenvectors
$r_{n,k}$ and $l_{n,k}$.
We define
\begin{equation}
    c_{n,k}(t)
    =
    \frac{
        l_{n,k}^\dagger
        \mathsf B_kq_k(t)
    }{
        l_{n,k}^\dagger
        \mathsf B_kr_{n,k}
    }.
    \label{eq:app_projection_coefficient}
\end{equation}

The corresponding contribution
to the boundary condensate is
\begin{equation}
    C_{n,k} = \left. \partial_z \left( r_{R,n,k} +ir_{I,n,k} \right) \right|_{z=0},\qquad \mathcal C_{n,k} =c_{n,k}C_{n,k}.
    \label{eq:app_projection_boundary}
\end{equation}
No factor of $1/2$ appears here,
because $q_k$ is already a
complex Fourier coefficient.

The actual boundary coefficient is
\begin{equation}
    \mathcal C_k^{\rm actual}
    =
    \left.
    \partial_z
    \left(\delta R+i\delta I\right)
    \right|_{z=0}.
\end{equation}
We define
\begin{equation}
    \eta_{n,k}
    =
    \frac{
        |\mathcal C_{n,k}|
    }{
        |\mathcal C_k^{\rm actual}|
    }.
    \label{eq:app_projection_ratio}
\end{equation}

This is a biorthogonal projection,
not an orthogonal decomposition.
The quantities $\eta_{n,k}$
are therefore not probabilities
or fractional energy contributions.

Table~\ref{tab:instantaneous_qnm_decomposition}
shows the tracked Landau-branch
contributions during the fastest
ramp, $R=10^{-3}$, for the
two modes involved in the
late-time reordering.

\begin{table}[t]
    \centering
    \small
    \caption{
        Instantaneous stationary-QNM
        decomposition of TD-driven
        perturbations at $R=10^{-3}$.
        Modes with $|\omega|<8$ are
        ranked by their individual
        boundary contributions.
        The quantity $\eta_{\rm L}$
        refers to the tracked
        Landau branch.
    }
    \label{tab:instantaneous_qnm_decomposition}
    \begin{tabular}{ccccc}
        \hline
        $v$
        & \shortstack{rank$_{\rm L}$\\$k=-0.750$}
        & \shortstack{$\eta_{\rm L}$\\$k=-0.750$}
        & \shortstack{rank$_{\rm L}$\\$k=-0.875$}
        & \shortstack{$\eta_{\rm L}$\\$k=-0.875$}
        \\
        \hline
        $0.450$ & $1$ & $1.1919$ & $1$ & $1.1238$ \\
        $0.470$ & $1$ & $1.4493$ & $1$ & $1.3089$ \\
        $0.490$ & $1$ & $1.9834$ & $1$ & $1.7238$ \\
        $0.500$ & $2$ & $2.3827$ & $2$ & $2.0545$ \\
        \hline
    \end{tabular}
\end{table}

The Landau branch gives the
largest individual boundary
contribution at $v=0.450$,
$0.470$ and $0.490$, but ranks
second at $v=0.500$ in both
sectors.

The values $\eta_{\rm L}>1$
reflect cancellations between
non-orthogonal modal contributions.
They do not imply that the
Landau mode carries more than
the total perturbation energy.

The decomposition is restricted
to QNMs with $|\omega|<8$.
No completeness of this truncated
set is assumed.

It is used only as a diagnostic.
The TD-driven and forced-background
calculations evolve the full
radial tangent equations and
do not depend on this QNM
truncation.

\section{Convergence and Robustness}
\label{app:robustness}

We collect the numerical checks supporting the
stationary QNM calculations, finite-rate mode
selection and history-dependent modal ordering.
We distinguish convergence under numerical
refinement from robustness under changes of
physical parameters.

\subsection{Radial spectral convergence}

We first test the stationary background and
QNM spectrum independently of the time evolution.

At
\begin{equation}
    \rho=10.0417,
    \qquad
    v=0.415,
\end{equation}
the counter-propagating sound branch is fitted
at small momentum using
\begin{equation}
    \operatorname{Re}\omega_-
    \simeq c_-|k|,
    \qquad
    \operatorname{Im}\omega_-
    \simeq\beta_-k^2.
    \label{eq:robust_sound_fit}
\end{equation}

Table~\ref{tab:qnm_radial_convergence}
shows the results at four radial resolutions.

\begin{table}[t]
    \centering
    \small
    \caption{
        Radial convergence of the stationary
        background and hydrodynamic QNM fit
        at $\rho=10.0417$ and $v=0.415$.
        The baseline resolution is $N_z=28$.
    }
    \label{tab:qnm_radial_convergence}
    \begin{tabular}{cccccc}
        \hline
        $N_z$
        & $\mu$
        & $c_-$
        & $c_+$
        & $\beta_-$
        & QNM residual
        \\
        \hline
        $20$
        & $6.5059600353$
        & $-0.0658340782$
        & $0.8424411501$
        & $0.0459137229$
        & $1.94\times10^{-16}$
        \\
        $24$
        & $6.5059600428$
        & $-0.0658335198$
        & $0.8424416967$
        & $0.0459148047$
        & $2.27\times10^{-16}$
        \\
        $28$
        & $6.5059600427$
        & $-0.0658335228$
        & $0.8424416938$
        & $0.0459147904$
        & $1.62\times10^{-16}$
        \\
        $32$
        & $6.5059600427$
        & $-0.0658335219$
        & $0.8424416948$
        & $0.0459147904$
        & $2.10\times10^{-16}$
        \\
        \hline
    \end{tabular}
\end{table}

Between $N_z=28$ and $32$, the changes in
$c_-$ and $c_+$ are of order $10^{-9}$,
while the change in $\beta_-$ is below
$10^{-10}$.

The Eddington--Finkelstein phase residual
decreases from $2.35\times10^{-8}$ at
$N_z=20$ to $2.84\times10^{-10}$ at
$N_z=24$, $2.97\times10^{-12}$ at
$N_z=28$ and $1.61\times10^{-13}$ at
$N_z=32$.

The QNM eigenproblem residual remains
at the $10^{-16}$ level, and the
nonlinear stationary background
residual remains below $4\times10^{-10}$
over the tested resolutions.

\subsection{Time-domain convergence and linearity}

The independent QNM--time-domain comparison
is given in Appendix~\ref{app:numerics}.
We further examine the frequency fitted
from direct evolution at
\begin{equation}
    \rho=10.0417,
    \qquad
    v=0.415,
    \qquad
    k=-0.500.
\end{equation}

The baseline calculation uses
\begin{equation}
    (N_z,N_x,\Delta t_{\max},\epsilon)
    =
    (28,128,0.020,10^{-5}).
\end{equation}

Table~\ref{tab:qnm_time_robustness}
gives the relative change in the
fitted frequency under numerical
refinement and changes of the
initial QNM amplitude. The
$N_x=256$ run also uses
$\Delta t_{\max}=0.01$, so it
is a combined refinement.

\begin{table}[t]
    \centering
    \small
    \caption{
        Robustness of the frequency
        extracted from direct evolution
        at $v=0.415$ and $k=-0.500$.
        Frequency errors are measured
        relative to the baseline
        time-domain fit.
    }
    \label{tab:qnm_time_robustness}
    \begin{tabular}{lccc}
        \hline
        Variation
        & \shortstack{Relative error in\\
          $\operatorname{Re}\omega$}
        & \shortstack{Relative error in\\
          $\operatorname{Im}\omega$}
        & \shortstack{Maximum\\constraint residual}
        \\
        \hline
        $\Delta t_{\max}=0.01$
        & $2.046\times10^{-10}$
        & $9.312\times10^{-12}$
        & $5.99\times10^{-12}$
        \\
        $N_x=256$,
        $\Delta t_{\max}=0.01$
        & $1.331\times10^{-10}$
        & $6.065\times10^{-11}$
        & $5.51\times10^{-12}$
        \\
        $\epsilon=10^{-6}$
        & $1.280\times10^{-10}$
        & $1.310\times10^{-9}$
        & $5.46\times10^{-12}$
        \\
        $\epsilon=10^{-4}$
        & $1.848\times10^{-7}$
        & $3.839\times10^{-7}$
        & $6.46\times10^{-12}$
        \\
        \hline
    \end{tabular}
\end{table}

The frequency is stable under the
tested refinements. Increasing the
initial amplitude from $10^{-5}$
to $10^{-4}$ changes both parts
of the fitted frequency by less
than $4\times10^{-7}$ relative
to the baseline.

We also test a second point,
\begin{equation}
    v=0.427,
    \qquad
    k=-0.500.
\end{equation}
The frequency-domain solver gives
\begin{equation}
    \omega_{\rm QNM}
    =
    -0.0460554406438
    +
    0.0138203581687\,i,
\end{equation}
while the time-domain fit gives
\begin{equation}
    \omega_{\rm evo}
    =
    -0.0460554408645
    +
    0.0138203583201\,i.
\end{equation}

The relative differences in the
real and imaginary parts are
$4.79\times10^{-9}$ and
$1.09\times10^{-8}$, respectively.

These tests establish agreement
between the independently computed
QNM frequencies and direct evolution.
They do not by themselves establish
the ordering of closely competing
threshold crossings.

\subsection{Eight-mode selection and crossing convergence}

We test the first-to-threshold
selection of Sec.~\ref{sec:selection}
using all eight seeded momenta,
\begin{equation}
    k=-\frac{m}{8},
    \qquad m=1,\ldots,8,
\end{equation}
with equal initial amplitudes
\begin{equation}
    A_k(0)=10^{-5}.
\end{equation}

For each mode, let $v_{\rm hit}(k)$
denote its first upward crossing
of the prescribed amplitude
threshold $A_{\rm th}$.
We define the separation between
the earliest and second-earliest
crossings as
\begin{equation}
    \delta v_{\rm hit}
    =
    v_{\rm hit}^{(2)}
    -
    v_{\rm hit}^{(1)}.
    \label{eq:app_winner_gap}
\end{equation}

We examine two cases with
closely competing modes.

The first uses
\begin{equation}
    R=4.0\times10^{-4},
    \qquad
    A_{\rm th}=10^{-4},
\end{equation}
where the leading candidates are
$k=-0.625$ and $-0.750$.

The second uses
\begin{equation}
    R=10^{-3},
    \qquad
    A_{\rm th}=10^{-3},
\end{equation}
where the full evolution selects
$k=-0.875$ rather than the
stationary-QNM transport prediction
$k=-0.750$.

Table~\ref{tab:eight_mode_selection_convergence}
compares the first crossing and
its separation from the runner-up
under numerical refinement.

\begin{table}[t]
    \centering
    \small
    \caption{
        Convergence of eight-mode
        first-to-threshold selection.
        The baseline resolution is
        $(N_z,N_x,\Delta t_{\max})
        =(28,128,0.020)$.
        The final column gives the
        separation between the first
        two crossings, obtained by
        linear interpolation of the
        recorded amplitudes.
    }
    \label{tab:eight_mode_selection_convergence}
    \begin{tabular}{lccc}
        \hline
        $(N_z,N_x,\Delta t_{\max})$
        & Winner
        & $v_{\rm hit}^{(1)}$
        & $\delta v_{\rm hit}$
        \\
        \hline
        \multicolumn{4}{c}{
            $R=4.0\times10^{-4}$,
            $A_{\rm th}=10^{-4}$
        }
        \\
        \hline
        $(28,128,0.020)$
        & $-0.625$
        & $0.454829640$
        & $2.856615\times10^{-6}$
        \\
        $(32,128,0.020)$
        & $-0.625$
        & $0.454829640$
        & $2.856615\times10^{-6}$
        \\
        $(28,128,0.010)$
        & $-0.625$
        & $0.454829640$
        & $2.856638\times10^{-6}$
        \\
        $(28,128,0.008)$
        & $-0.625$
        & $0.454829640$
        & $2.846923\times10^{-6}$
        \\
        $(28,256,0.008)$
        & $-0.625$
        & $0.454829640$
        & $2.846925\times10^{-6}$
        \\
        \hline
        \multicolumn{4}{c}{
            $R=10^{-3}$,
            $A_{\rm th}=10^{-3}$
        }
        \\
        \hline
        $(28,128,0.020)$
        & $-0.875$
        & $0.497088981$
        & $6.792642\times10^{-5}$
        \\
        $(32,128,0.020)$
        & $-0.875$
        & $0.497088981$
        & $6.792643\times10^{-5}$
        \\
        $(28,128,0.010)$
        & $-0.875$
        & $0.497088981$
        & $6.792644\times10^{-5}$
        \\
        $(28,128,0.008)$
        & $-0.875$
        & $0.497088860$
        & $6.806717\times10^{-5}$
        \\
        $(28,256,0.008)$
        & $-0.875$
        & $0.497088860$
        & $6.806717\times10^{-5}$
        \\
        \hline
    \end{tabular}
\end{table}

The winner is unchanged in
every tested calculation.
For the slower ramp, the first
two crossings are separated by
approximately $2.86\times10^{-6}$.
The separation remains positive
under radial, temporal and
Fourier refinement.

For the fastest ramp, the
baseline crossings are
\begin{equation}
    v_{\rm hit}(-0.875) = 0.4970889807,\qquad v_{\rm hit}(-0.750) = 0.4971569071.
    \label{eq:app_fast_hit_values}
\end{equation}
Their separation is approximately
$6.79\times10^{-5}$.

We also test the interpolation
used to locate the threshold
crossings. For the fastest
ramp, linear interpolation
gives
\begin{equation}
    \delta v_{\rm hit}^{\rm linear}
    =
    6.79264\times10^{-5},
\end{equation}
interpolation of the logarithmic
amplitude gives
\begin{equation}
    \delta v_{\rm hit}^{\log}
    =
    6.80236\times10^{-5},
\end{equation}
and piecewise cubic Hermite
interpolation gives
\begin{equation}
    \delta v_{\rm hit}^{\rm PCHIP}
    =
    6.80351\times10^{-5}.
    \label{eq:app_crossing_interpolation}
\end{equation}

All three methods preserve
the first-to-second ordering.
The same interpolation check
preserves the winner in the
$R=4.0\times10^{-4}$ case.

The runs with
$\Delta t_{\max}=0.008$
require a qualification.
Their Fourier-filtering cadence
was adjusted in step count to
keep approximately the same
physical filtering interval.
The recorded output intervals
also differ slightly from those
of the other runs. Consequently,
the small changes in the
linearly interpolated crossing
positions cannot be assigned
solely to the time integrator.

The comparisons between
$\Delta t_{\max}=0.020$ and
$0.010$, and between
$N_x=128$ and $256$ at fixed
$\Delta t_{\max}=0.008$,
provide separate checks of
time-step and Fourier-resolution
dependence. Together with the
interpolation tests, they show
that the selected modes are
numerically stable.

The winner also remains unchanged
under these refinements at the
other tested thresholds in the
two ramp-rate suites.

\subsection{Time-dependent linear evolution}

The finite-momentum tangent
equations are checked against
the frequency-domain QNM
eigenproblem, as described in
Appendix~\ref{app:td_linear}.

For the six tested momentum
sectors,
\begin{equation}
    k=-\frac{m}{8},
    \qquad m=3,\ldots,8,
\end{equation}
the largest normalised
QNM--tangent residual is
$1.70\times10^{-10}$.
The largest mismatch in the
reconstructed temporal gauge
field is $3.21\times10^{-17}$.

We also compare the
time-dependent linear evolution
on the driven background with
the complete eight-mode PDE
evolution.

At $R=10^{-3}$, the ratio
$A_{-0.875}/A_{-0.750}$
crosses unity at
\begin{equation}
    v_\times^{\rm full}
    =
    0.49472528,
    \qquad
    v_\times^{\rm TD\text{-}driven}
    =
    0.49472880.
    \label{eq:app_td_crossing_check}
\end{equation}
The difference is approximately
$3.52\times10^{-6}$.

This agreement supports the
time-dependent linear description
of the competing modes in the
regime considered. It is a
comparison of two evolution
models, rather than a numerical
convergence test of either model.

In particular, it does not
replace an independent
convergence test of the
forced-background reconstruction
introduced in
Sec.~\ref{sec:tdlinear}.

\subsection{Convergence of the memory reversal}

We examine the history-dependent
reversal of Sec.~\ref{sec:memory}
at
\begin{equation}
    v_i=0.395,\qquad \dot v=4.25\times10^{-4},\qquad v_h=0.4086,\qquad v_{\rm obs}=0.460,
\end{equation}
for the pair
\begin{equation}
    (k_i,k_j)=(-0.625,-0.750).
\end{equation}

The driving protocol and initial
perturbations are held fixed
while the numerical resolution
is varied.

Define
\begin{equation}
    \Delta_{20}
    =
    \Delta_{\rm mem}(20),
    \qquad
    \Delta_{30}
    =
    \Delta_{\rm mem}(30).
\end{equation}
Since these values bracket the
reversal, we estimate its
position by
\begin{equation}
    \tau_h^*
    =
    20-
    10\frac{\Delta_{20}}
    {\Delta_{30}-\Delta_{20}}.
    \label{eq:tau_star_interpolation}
\end{equation}

Table~\ref{tab:memory_numerical_convergence}
gives the results.

\begin{table}[t]
    \centering
    \small
    \caption{
        Convergence of the
        history-dependent pairwise
        reversal.
        The last column gives the
        relative displacement of
        the interpolated reversal
        time from the baseline.
    }
    \label{tab:memory_numerical_convergence}
    \begin{tabular}{ccccc}
        \hline
        $(N_z,N_x,\Delta t_{\max})$
        & $\Delta_{20}$
        & $\Delta_{30}$
        & $\tau_h^*$
        & \shortstack{Relative\\change}
        \\
        \hline
        $(28,128,0.020)$
        & $+7.6354717\times10^{-3}$
        & $-9.5820121\times10^{-3}$
        & $24.43471986$
        & $0$
        \\
        $(28,256,0.020)$
        & $+7.6356233\times10^{-3}$
        & $-9.5818630\times10^{-3}$
        & $24.43480724$
        & $3.576\times10^{-6}$
        \\
        $(40,128,0.020)$
        & $+7.6354696\times10^{-3}$
        & $-9.5820121\times10^{-3}$
        & $24.43471916$
        & $2.851\times10^{-8}$
        \\
        $(28,128,0.010)$
        & $+7.6359206\times10^{-3}$
        & $-9.5815624\times10^{-3}$
        & $24.43498077$
        & $1.068\times10^{-5}$
        \\
        \hline
    \end{tabular}
\end{table}

The signs
\begin{equation}
    \Delta_{20}>0,
    \qquad
    \Delta_{30}<0
\end{equation}
are preserved under every
refinement. The largest relative
change in the interpolated
reversal time is
$1.07\times10^{-5}$.

The numerical results therefore
support the stability of the
pairwise reversal.

\subsection{Constraint and conservation diagnostics}

For the nonlinear evolutions,
we monitor the radial Maxwell
constraint, conservation of the
spatially averaged charge density,
and the imposed boundary conditions.

We define
\begin{equation}
    \epsilon_{\rm C} = \max_t \|\mathcal C_{\rm M}(t)\|_\infty,\qquad \epsilon_Q = \max_t \frac{ |\bar\rho(t)-\bar\rho(0)| }{ \bar\rho(0) },\qquad \epsilon_{\rm B} = \max_t \|\mathcal B(t)\|_\infty,
    \label{eq:numerical_health_measures}
\end{equation}
where $\mathcal C_{\rm M}$
is the radial Maxwell
constraint residual and
$\mathcal B$ denotes the
boundary-condition errors.

Table~\ref{tab:numerical_health_robustness}
gives the maximum values
recorded in the principal
numerical suites.

\begin{table}[t]
    \centering
    \small
    \caption{
        Numerical diagnostics for
        the eight-mode convergence
        runs, memory convergence
        tests and two physical
        generalisation studies.
        Values are maxima over
        the indicated runs.
    }
    \label{tab:numerical_health_robustness}
    \begin{tabular}{lccc}
        \hline
        Calculation
        & $\epsilon_{\rm C}$
        & $\epsilon_Q$
        & $\epsilon_{\rm B}$
        \\
        \hline
        Eight-mode convergence
        & $9.39\times10^{-12}$
        & $7.08\times10^{-16}$
        & $1.95\times10^{-14}$
        \\
        Memory convergence
        & $1.68\times10^{-11}$
        & $5.31\times10^{-16}$
        & $2.31\times10^{-14}$
        \\
        Second background, $\rho=11.5$
        & $6.56\times10^{-12}$
        & $1.54\times10^{-16}$
        & $1.95\times10^{-14}$
        \\
        Second pair, $(-0.750,-0.875)$
        & $6.14\times10^{-12}$
        & $1.77\times10^{-16}$
        & $1.78\times10^{-14}$
        \\
        \hline
    \end{tabular}
\end{table}

The constraint residuals,
charge drift and boundary errors
remain small throughout these
calculations.

The second-background and
second-pair runs change the
physical system and driving
protocol. They test the
robustness of the pairwise
memory effect rather than
numerical convergence.

Likewise, the gain-law errors
in Table~\ref{tab:memory_generalisation}
measure the accuracy of a
reduced physical approximation,
not the accuracy of the
numerical discretisation.

The convergence tests support
the stationary QNM results,
the eight-mode threshold
ordering and the
history-dependent reversal.
The consistency between
Full PDE and TD-driven
provides a separate check
of the time-dependent
linear description.

\clearpage
\bibliography{Ref}

\end{document}